\documentclass{aa}  

\usepackage{graphicx}
\usepackage{txfonts}
\usepackage{enumitem}
\usepackage[utf8]{inputenc} 
\usepackage{cancel}
\usepackage{graphicx}
\usepackage{xcolor}
\usepackage{geometry}
\usepackage[fleqn]{amsmath}
\usepackage{mathtools}
\usepackage{epstopdf}
\usepackage{multirow}
\usepackage{txfonts}
\usepackage{expdlist}
\usepackage{natbib}
\usepackage{color}
\usepackage{epsfig}
\usepackage{pdflscape}
\usepackage{lscape}
\usepackage{rotating}
\usepackage{url}
\usepackage{natbib,twoopt}
\usepackage[hyphenbreaks]{breakurl}
\usepackage[breaklinks]{hyperref}  
\DeclareMathAlphabet\mathzapf       {T1}{pzc} {mb} {it}
\bibpunct{(}{)}{;}{a}{}{,}             
\definecolor{cobalt}{rgb}{0.06, 0.2, 0.65}
\hypersetup{
  colorlinks,
  citecolor=cobalt,
  linkcolor=cobalt,
  urlcolor=cobalt,
}
\makeatletter
  \newcommandtwoopt{\citeads}[3][][]{\href{http://adsabs.harvard.edu/abs/#3}%
    {\def\hyper@linkstart##1##2{}%
     \let\hyper@linkend\@empty\citealp[#1][#2]{#3}}}
  \newcommandtwoopt{\citepads}[3][][]{\href{http://adsabs.harvard.edu/abs/#3}%
    {\def\hyper@linkstart##1##2{}%
     \let\hyper@linkend\@empty\citep[#1][#2]{#3}}}
  \newcommandtwoopt{\citetads}[3][][]{\href{http://adsabs.harvard.edu/abs/#3}%
    {\def\hyper@linkstart##1##2{}%
     \let\hyper@linkend\@empty\citet[#1][#2]{#3}}}
  \newcommandtwoopt{\citeyearads}[3][][]%
    {\href{http://adsabs.harvard.edu/abs/#3}
    {\def\hyper@linkstart##1##2{}%
     \let\hyper@linkend\@empty\citeyear[#1][#2]{#3}}}
\makeatother
\newcommand{\grau}      {$^{\circ}$}

\newcommand{\mJy}               {mJy~beam$^{-1}$}
\newcommand{\Jy}                {Jy~beam$^{-1}$}

\def\arc{\mbox{$^{\prime\prime}$}}

\def\deg{\mbox{$^{\circ}$}}

\def\flux {\mbox{erg~cm$^{-2}$~s$^{-1}$}}
\def\lum {\mbox{erg~s$^{-1}$}}

\def\xmm{{\it XMM-Newton}}

\newcommand{\cxo}{{\em Chandra}}
\newcommand{\swift}{{\em Swift}}

\newcommand{\nustar}{{\em NuSTAR}}

\def\srcfirst {\mbox{\object{PSR\,J1023$+$0038}}}
\def\src {\mbox{J1023}}
\def\srcmm {\mbox{ALMA\,J102349.1$+$003824}}

\DeclareUnicodeCharacter{2212}{\ensuremath{-}}

\begin{document} 

\title{Matter ejections behind the highs and lows of the transitional millisecond pulsar \srcfirst}

 %  \subtitle{ }

\author{M.~C.~Baglio\inst{1,2}
          \and
          F.~Coti~Zelati\inst{3,4,2}\fnmsep\thanks{The first two authors equally contributed to this work.}
          \and
          S.~Campana\inst{2}
          \and
          G.~Busquet\inst{5,6,4}
          \and
          P.~D'Avanzo\inst{2}
          \and
          S.~Giarratana\inst{7,8}
          \and
          M.~Giroletti\inst{7}
          \and
          F.~Ambrosino\inst{9,10,11}
          \and
          S.~Crespi\inst{1}
          \and
          A.~Miraval Zanon\inst{12,9}
          \and
          X.~Hou\inst{13,14}
          \and
          D.~Li\inst{15,16,17}
          \and
          J.~Li\inst{18,19}
          \and
          P.~Wang\inst{15, 20}
          \and
          D.~M.~Russell\inst{1}
          \and
          D.~F.~Torres\inst{3,4,21}
          \and
          K.~Alabarta\inst{1}
          \and
          P.~Casella\inst{9}
          \and
          S.~Covino\inst{2}
          \and
          D.~M.~Bramich\inst{1,22}
          \and
          D.~de~Martino\inst{23}
          \and
          M.~M\'{e}ndez\inst{24}
          \and
          S.~E.~Motta\inst{2}
          \and
          A.~Papitto\inst{9}
          \and
          P.~Saikia\inst{1}
          \and
          F.~Vincentelli\inst{25,26}
          }

\institute{Center for Astro, Particle, and Planetary Physics, New York University Abu Dhabi, P.O. Box 129188, Abu Dhabi, UAE\\
            \email{mcb19@nyu.edu}
            \and
            INAF--Osservatorio Astronomico di Brera, Via Bianchi 46, I-23807 Merate (LC), Italy
        \and
            Institute of Space Sciences (ICE, CSIC), Campus UAB, Carrer de Can Magrans s/n, E-08193 Barcelona, Spain\\
            \email{cotizelati@ice.csic.es}
            \and
            Institut d'Estudis Espacials de Catalunya (IEEC), Carrer Gran Capit\`a 2--4, E-08034 Barcelona, Spain
            \and
            Departament de F\'isica Qu\`antica i Astrof\'isica, Universitat de Barcelona (UB), c/ Mart\'i Franqu\`es 1, E-08028 Barcelona, Spain
            \and
            Institut de Ci\`encies del Cosmos (ICCUB), Universitat de Barcelona (UB), c/ Mart\'i Franqu\`es 1, E-08028 Barcelona, Spain
           \and
           INAF--Istituto di Radioastronomia, via Gobetti 101, I-40129, Bologna, Italy
           \and
           Department of Physics and Astronomy, University of Bologna, via Gobetti 93/2, I-40129 Bologna, Italy
           \and
           INAF--Osservatorio Astronomico di Roma, Via Frascati 33, I-00078 Monteporzio Catone (RM), Italy
           \and
           INAF--Istituto di Astrofisica e Planetologia Spaziali, Via Fosso del Cavaliere 100, I-00133 Rome, Italy 
           \and
           Sapienza Università di Roma, Piazzale Aldo Moro 5, I-00185 Rome, Italy
           \and
           ASI - Agenzia Spaziale Italiana, Via del Politecnico snc, I-00133 Rome, Italy
           \and
           Yunnan Observatories, Chinese Academy of Sciences, 650216 Kunming, China
           \and
           Key Laboratory for the Structure and Evolution of Celestial Objects, Chinese Academy of Sciences, Kunming 650216, China
           \and
           National Astronomical Observatories, Chinese Academy of Sciences, 20A Datun Road, Chaoyang District, Beijing, 100101, China
           \and
           University of Chinese Academy of Sciences, Beijing, 100049, China  
           \and
           Research Center for Intelligent Computing Platforms, Zhejiang Laboratory, Hangzhou, 311100, China
           \and
           CAS Key Laboratory for Research in Galaxies and Cosmology, Department of Astronomy, University of Science and Technology of China, Hefei, China
           \and
           School of Astronomy and Space Science, University of Science and Technology of China, Hefei, China
           \and
           Institute for Frontiers in Astronomy and Astrophysics, Beijing Normal University, Beijing, 102206, China
           \and
           Institució Catalana de Recerca i Estudis Avançats (ICREA), Passeig Llu\'is Companys 23, E-08010 Barcelona, Spain
           \and
           Division of Engineering, New York University Abu Dhabi, P.O. Box 129188, Saadiyat Island, Abu Dhabi, UAE
           \and
           INAF−-Osservatorio Astronomico di Capodimonte, Salita Moiariello 16, I-80131 Napoli, Italy
           \and           
           Kapteyn Astronomical Institute, University of Groningen, PO BOX 800, NL-9700 AV Groningen, The Netherlands
           \and  
           Instituto de Astrof\'{i}sica de Canarias (IAC), V\'ia L\'actea s/n, E-38205 La Laguna, S/C de Tenerife, Spain
           \and  
           Departamento de Astrof\'{i}sica, Universidad de La Laguna, Avda. Astrof\'isico Francisco S\'anchez s/n, E-38206 La Laguna, S/C de Tenerife, Spain
           }   

\date{Received 14 March 2023 / Accepted 17 July 2023}

% \abstract{}{}{}{}{} 
% 5 {} token are mandatory
 
\abstract{
Transitional millisecond pulsars are an emerging class of sources that link low-mass X-ray binaries to millisecond radio pulsars in binary systems. These pulsars alternate between a radio pulsar state and an active low-luminosity X-ray disc state. During the active state, these sources exhibit two distinct emission modes (high and low) that alternate unpredictably, abruptly, and incessantly.
X-ray to optical pulsations are observed only during the high mode. 
The root cause of this puzzling behaviour remains elusive. 
This paper presents the results of the most extensive multi-wavelength campaign ever conducted on the transitional pulsar prototype, PSR\,J1023$+$0038, covering from the radio to X-rays. The campaign was carried out over two nights in June 2021 and involved 12 different telescopes and instruments, including \emph{XMM-Newton}, HST, VLT/FORS2 (in polarimetric mode), ALMA, VLA, and FAST.
By modelling the broadband spectral energy distributions in both emission modes, we show that the mode switches are caused by changes in the innermost region of the accretion disc. These changes trigger the emission of discrete mass ejections, which occur on top of a compact jet, as testified by the detection of at least one short-duration millimetre flare with ALMA at the high-to-low mode switch. The pulsar is subsequently re-enshrouded, completing our picture of  the mode switches.}


\titlerunning{Matter ejections behind the highs and lows of PSR J1023+0038}
\authorrunning{Baglio~M.~C., Coti Zelati~F. et al.}
\maketitle
%
%-------------------------------------------------------------------

\section{Introduction}
The evolutionary path that leads a neutron star (NS) in a low-mass X-ray binary (XRB) system to become a millisecond radio pulsar involves different stages, the most enigmatic of which is represented by transitional millisecond pulsars (tMSPs). These sources alternate between a radio pulsar state and an active low-luminosity X-ray disc state on timescales from weeks to months (for reviews, see \citealt{campana18b,papitto22}).
The archetypal tMSP is PSR\,J1023$+$0038 (hereafter referred to as \src), which was discovered in 2007 as a 1.69\,ms radio pulsar orbiting a low-mass ($\simeq$0.2\,$M_\odot$) companion star with a period of 4.75\,hr \citep{Archibald2009}. 
In 2013, \src\ showed a sudden brightening in emission at X-ray and gamma-ray frequencies (by a factor of 5--10) as well as at ultraviolet (UV) and optical frequencies (by 1--2\,mag), which coincided with the disappearance of the pulsed radio signal \citep{stappers14,patruno14}. Double-peaked optical emission lines were soon detected, indicating the formation of an accretion disc \citep{cotizelati14}.  \src\ has remained in this active state ever since, at an X-ray luminosity of $L_X\simeq 7\times 10^{33}$\,erg\,s$^{-1}$ in the 0.3--80\,keV energy range \citep{cotizelati18a} for a distance of 1.37\,kpc \citep{deller12}. 

During the sub-luminous X-ray state, the X-ray emission of \src\ is observed to switch between two distinct intensity modes, dubbed high and low, with sporadic flares in between (e.g. \citealt{bogdanov15}). The high mode occurs 70--80\% of the time, while the low mode occurs 20--30\% of the time and typically lasts from a few tens of seconds to minutes. The drop and rise timescales involved in the mode switching are of the order of $\approx$10\,s. 
X-ray, UV, and optical pulsations occur simultaneously in the X-ray high-intensity mode and disappear in the low mode \citep{archibald15,papitto19,miravalzanon22,illiano23}. The UV, optical, and near-infrared (NIR) emission is dominated by the accretion disc and irradiated companion star and also shows flaring and flickering activities \citep{shahbaz15,kennedy18,papitto18,hakala18,shahbaz18}. 
Optical polarimetric observations reveal a linear polarisation (LP) at a level of $\sim$1\% \citep{baglio16,hakala18} of uncertain origin. 
Bright, variable radio continuum emission is also detected \citep{deller15}. In particular, the low X-ray mode episodes are accompanied by an increase in the radio flux by a factor of $\simeq$3 on average, with fairly equal rise and decay times and an overall duration that matches low X-ray mode intervals \citep[][see also our Fig.\,\ref{fig:timeseries1}]{bogdanov18}. These re-brightenings are associated with a temporal evolution of the radio spectrum from inverted to steep. 
Currently, \src\ spins down at a rate that is $\simeq$4\% faster than in the radio pulsar state \citep{burtovoi20}. This overall behaviour is unique in the landscape of XRBs, and no comprehensive explanation for it has been found.

This paper presents the results of the largest multi-wavelength campaign ever performed on \src. Our aim is to ultimately understand the root cause of the mode changes in the active sub-luminous X-ray state and its implication for our understanding of the accretion-ejection coupling in XRB systems that harbour rapidly spinning NSs.
The observations of our campaign were conducted on two different nights in June 2021. The first night (Night\,1) involved \xmm, the Nuclear Spectroscopic Telescope Array (\nustar), the \textit{Hubble} Space Telescope (HST), the Son OF ISAAC (SOFI) instrument mounted on the New Technology Telescope (NTT), the \textit{Karl G. Jansky} Very Large Array (VLA), and the Five-hundred-meter Aperture Spherical Telescope (FAST); the second night (Night\,2) involved the Neutron Star Interior Composition Explorer Mission (NICER), the \textit{Neil Gehrels Swift} Observatory (\swift), the Very Large Telescope (VLT), the Rapid Eye Mount (REM) telescope, and the Atacama Large millimetre/sub-millimetre Array (ALMA).

The paper is structured as follows: 
in Sect.\,\ref{sec:obs} we present the observations and data analysis. The results are then reported in Sect.\,\ref{sec:results}. In Sect.\,\ref{sec:discussion} we propose a scenario for the mode switching, present the results of the modelling of the spectral energy distributions (SEDs) in the two modes, and discuss the results in relation to our scenario. Conclusions follow in Sect.\,\ref{sec:conclusions}.

%%%%%%%%%%%%%%%%%%%%%%%%%%%%%%%%%%%%%%%%%%%%%%%%%%%%%%%%%%%%%%%%%%%%%%%%
\begin{table*}
\small
\caption{
\label{tab:log}
Observation log of \src\ in June 2021. Observations are listed in order of decreasing observing frequency. } 
\centering
\begin{tabular}{lcccc}
\hline\hline
Telescope/Instrument    & Obs. ID/Project       & Start -- End time                       & Exposure  & Band (setup) \\
                                      &                       & MMM DD hh:mm:ss (UTC)                   & (ks)      &   \\
\hline
\multicolumn{5}{c}{Night\,1}\\
\hline
\xmm/EPIC-pn            & 0864010101        & Jun 3 20:44:32  -- Jun 4 13:39:15   & 58.7      & 0.3--10\,keV (fast timing)   \\
\xmm/EPIC-MOS1          & 0864010101        & Jun 3 21:09:37  -- Jun 4 13:39:01   & 58.1      & 0.3--10\,keV (full frame)  \\
\xmm/EPIC-MOS2          & 0864010101        & Jun 3 22:31:54  -- Jun 4 13:39:02   & 53.5      & 0.3--10\,keV (full frame)  \\
\nustar/FPMA/FPMB       & 30601005002       & Jun 3 20:36:09  -- Jun 4 10:06:09   & 22.3/22.1 & 3--79\,keV   \\
\swift/XRT              & 00033012207       & Jun 3 18:29:46 --  Jun 3 23:45:44   & 6.4       & 0.3--10\,keV (photon counting)  \\
\swift/XRT              & 00033012208       & Jun 4 02:27:57 --  Jun 4 12:17:45   & 9.5       & 0.3--10\,keV (photon counting)   \\
\swift/XRT              & 00033012209       & Jun 4 03:50:46 --  Jun 4 08:44:56   & 2.0       & 0.3--10 keV (photon counting)   \\
\swift/UVOT             & 00033012207       & Jun 3 18:29:50  -- Jun 3 23:45:46   & 6.3       & UVM2 (Event)   \\
\swift/UVOT             & 00033012209       & Jun 4 03:50:50  -- Jun 4 08:44:58   & 1.9       & UVM2 (Event)  \\
HST STIS/NUV-MAMA & 16061             & Jun 4 00:09:07  -- Jun 4 00:43:27   & 2.1       & G230L (fast timing spectroscopy)   \\
HST STIS/NUV-MAMA & 16061             & Jun 4 01:35:28  -- Jun 4 02:20:18   & 2.7       & G230L (fast timing spectroscopy)   \\
\xmm/OM                 & 0864010101        & Jun 3 22:51:01  -- Jun 4 13:23:52   & 48.4      & $B$ (fast window)  \\
NTT/SOFI                    &                   & Jun 3 23:49:09 -- Jun 4 02:42:38    & 10.4      & $J$ (fast photometry)   \\
VLA                     & SJ6401            & Jun 3 22:30:00  -- Jun 4 02:29:20   & 6.7       & Band $X$ (C configuration)  \\
FAST                    & PT2020$_-$0044    & Jun 4 08:47:45 -- Jun 4 11:47:45    & 10.8      & Band $L$  \\
\hline
\multicolumn{5}{c}{Night\,2}\\
\hline
NICER/XTI             & 4034060102   & Jun 26 21:56:04 -- Jun 26 22:34:31 & 2.0       & 0.3--10\,keV   \\
NICER/XTI             & 4034060103   & Jun 26 23:29:04 -- Jun 27 01:33:51 & 4.1       & 0.3--10\,keV   \\
\swift/XRT                      & 00033012214  & Jun 26 22:19:35 -- Jun 26 22:39:47 & 1.2       & 0.3--10\,keV (photon counting)  \\
\swift/XRT                      & 00033012213  & Jun 26 23:56:12 -- Jun 26 23:58:46 & 0.15      & 0.3--10\,keV (photon counting)  \\
\swift/XRT                      & 00033012215  & Jun 27 00:04:44 -- Jun 27 00:21:47 & 1.0       & 0.3--10\,keV (photon counting)  \\
\swift/UVOT                 & 00033012214  & Jun 26 22:19:35 -- Jun 26 22:39:47 & 1.2       & UVM2 (Event+Imaging)  \\
\swift/UVOT             & 00033012213  & Jun 26 23:56:12 -- Jun 26 23:58:46 & 0.15      & UVM2 (Event+Imaging)  \\
\swift/UVOT                 & 00033012215  & Jun 27 00:04:44 -- Jun 27 00:21:47 & 1.0       & UVM2 (Event+Imaging)   \\
VLT/FORS2               & 107.22RK.001 & Jun 26 22:51:56 -- Jun 27 01:15:30 & 8.7       & $R $ \\
REM/ROSS2               &   43013      & Jun 26 23:20:33 -- Jun 27 01:24:41 & 7.4       & $griz$\\
REM/REMIR               &   43013      & Jun 26 23:20:44 -- Jun 27 01:25:34 & 7.4       & $K$ \\
ALMA                        & 2019.A.00036.S & Jun 26 21:58:42 -- Jun 27 01:12:06 & 9.3     & Band 3 (C43-7 configuration)   \\
\hline
\end{tabular}
\end{table*}
%%%%%%%%%%%%%%%%%%%%%%%%%%%%%%%%%%%%%%%%%%%%%%%%%%%%%%%%%%%%%%%%%%%%%%%%

\section{Observations and data analysis}
\label{sec:obs}
Table\,\ref{tab:log} lists all the observations of \src\ presented in this paper. In the following, we describe these observations and the procedures adopted to process and analyse these data.

\subsection{\xmm\ (Night\,1)}
\xmm\ \citep{jansen01,schartel22} observed \src\ on 2021 June 3--4 (Obs. ID 0864010101) using the European Photon Imaging Cameras (EPIC) and the Optical/UV Monitor (OM) telescope \citep{mason01}. The EPIC-pn \citep{struder01} was set in fast timing mode (time resolution of 29.52\,$\mu$s), the two EPIC-MOS (Metal Oxide Semi-conductor; \citealt{turner01}) in full frame mode (time resolution of 2.6\,s) and the OM in the fast window mode (time resolution of 0.5\,s) with the $B$ filter (effective wavelength 4330\,\AA; full width at half maximum 926\,\AA). The observation data files were processed and analysed using the Science Analysis System (SAS; v. 20.0). The X-ray data were analysed in the same way as described by \cite{miravalzanon22}. No periods of high background activity were detected.
The background-subtracted light curve was extracted over the time interval during which all three EPIC instruments collected data simultaneously and was binned with a time resolution of 10\,s. 
The optical background-subtracted light curve was extracted using the \textsc{omfchain} pipeline with default parameters. The light curve was binned at a time resolution of 120\,s to ensure a high S/N and detect any changes in the optical brightness at least during prolonged periods of low X-ray mode. The net count rates were then converted to magnitudes in the Vega system using the formula provided on the \xmm\ online threads\footnote{\url{https://www.cosmos.esa.int/web/xmm-newton/sas-watchout-uvflux}}.
To align the X-ray and optical light curves with those extracted from other observatories, the arrival times of photons were converted from the terrestrial time standard to the Coordinated Universal Time (UTC) standard without applying a barycentric correction\footnote{This same procedure was also applied to the data collected with \nustar, \swift\ and NICER presented in the following.}.

\subsection{\nustar\ (Night\,1)}
\nustar\ \citep{harrison13} observed \src\ for a total of $\sim$48.6\,ks and a net exposure of $\sim$22\,ks (Obs. ID 30601005002), almost fully overlapping with the \xmm\ observation. Data were processed and analysed using the NuSTAR Data Analysis Software (NUSTARDAS, v.2.1.2) with the instrumental calibration files stored in CALDB v20220525 and the same prescriptions reported by \cite{miravalzanon22}. \src\ was detected up to energies of $\sim$50\,keV in both focal plane modules (FPMA and FPMB). Light curves were extracted separately for the two modules, combined to increase the S/N, and binned at a time resolution of 50\,s. 

\subsection{HST (Night\,1)}
The Space Telescope Imaging Spectrograph (STIS; \citealt{woodgate98}) on board HST observed \src\ for 4750\,s using the Near-UV Multi-Anode Micro-channel Array  (NUV-MAMA) detector in TIME-TAG mode (time resolution of 125\,$\mu$s); the data were collected with the G230L grating equipped with a 52\arc\ $\times$ 0.2 slit with a spectral resolution of $\sim$500 over the nominal range (first order). 
The data analysis procedure that we adopted is the same as that described by  \cite{miravalzanon22}. In summary, we used the \texttt{stis$\_$photons} package\footnote{\url{https://github.com/Alymantara/stis_photons}} to correct the position of the slit channels and to assign the wavelengths to each         time of arrival. We selected times of arrival  belonging to channels 993--1005 of the slit and in the 165--310\,nm wavelength interval to isolate the source signal.

\subsection{NTT/SOFI (Night\,1)}
We collected NIR high time-resolution imaging data using the SOFI \citep{moorwood98} instrument mounted on the European Southern Observatory (ESO) 3.6 m NTT at La Silla Observatory (Chile). Observations were carried out using the $J$ filter from 2021 June 3 at 23:49:09 UTC to June 4 at 02:42:38 UTC (Table\,\ref{tab:log}). In order to achieve 1 *s time resolution, observations were performed in Fast Photometry mode, which restricts the area on which the detector is read. Specifically, the instrument was set to read a 400$\times$400 pixels window ($1.9'\times1.9'$). During the observation, a series of 100 frames each with an exposure time of 1.0\,s were stacked together to form a `data cube'. There was a gap of $\sim$3\,s between each cube. Photometric data were extracted through aperture photometry with an aperture radius of $1.4''$ using a custom pipeline based on the \texttt{DAOPHOT} package \citep{DAOPHOT}. In order to minimise systematic errors, we performed photometry on \src\ and a reference star (at position R.A. = 10$^\mathrm{h}$23$^\mathrm{m}$50$^\mathrm{s}$54, Dec. = +00$^{\circ}$38$^{\prime}$16.$^{\prime\prime}$4, J2000.0, and with a magnitude of $J = 15.44 \pm 0.05$ mag in the Vega system from the Two Micron All-Sky Survey (2MASS) point source catalogue), assumed to have constant brightness. We then re-binned the light curve at a time resolution of 10\,s to reduce the scatter and investigate the presence of variability at the same time resolution as the light curves in the X-ray and UV bands. For a more conservative approach, we evaluated the errors on the magnitudes of the re-binned light curves as the standard deviation of the magnitudes measured in each bin.

\subsection{VLA (Night\,1)}
The VLA observed \src\ for 4 hours, starting on 2021 June 3 at 22:30 UTC. The array was in its C-configuration and data were taken at a central frequency of 10\,GHz, with a bandwidth of 4\,GHz. J1331$+$3030 was used as flux and band-pass calibrator, while J1024$-$0052 was used as phase calibrator. The separation between the target and the phase calibrator is 0.3$^{\circ}$. To calibrate and analyse the data we followed the prescription of \cite{deller15} and \cite{bogdanov18}. Here we report the main steps. After initial calibration using the VLA pipeline within the Common Astronomy Software Application (\textsc{casa}) package (version 5.1.1), manual flagging of bad data was performed. Unfortunately, the first half of the observation turned out to be unusable, possibly due to the bad weather conditions that occurred during the observation, and was completely flagged out. 

Subsequently, we self-calibrated the data using a solution interval of 75\,s for the phase calibration and 45\,min for the amplitude$+$phase calibration. Upon inspection of the field of view (FOV), three main sources were identified: the target \src, a faint source to the north, namely J102348.2$+$004017, and the brightest source in the field, located to the south-east, namely J102358.2$+$003826. J102358.2$+$003826 is a well-known galaxy at redshift $z = 0.449$ \citep{ahn14} with a resolved extended emission whose secondary lobes heavily affect the overall root mean square (rms) noise level \citep{deller15}. To remove the problematic contribution of J102358.2$+$003826, we first cleaned the image using a mask with the \texttt{tclean} task. We set the \texttt{nterms} parameter equal to 2 to take into account the spectral properties of the galaxy over the wide bandwidth of the observation \citep{deller15}. This step provided the model visibilities of J102358.2$+$003826. Next, we used the \texttt{uvsub} task to subtract these model visibilities from the self-calibrated visibilities and deconvolved the residuals. The resulting image therefore does not have any contribution from J102358.2$+$003826. 

As pointed out by \cite{deller15}, the gain solutions from the self-calibration obtained before the subtraction of J102358.2$+$003826 were dominated by the pointing errors of this source. This source was located beyond the half-power point of the antenna, resulting in inaccurate solutions. Therefore, we had to invert the amplitude and phase self-calibration after subtracting this source from the field. We used the \textsc{casa} task \texttt{invgain} \citep{hales16} to invert the calibration table and applied it to the self-calibrated visibilities. The final image obtained had a beamwidth size of 3.6$^{\prime\prime} \times$2.1$^{\prime\prime}$, and an rms noise level of $\simeq$4\,$\mu$Jy\,beam$^{-1}$. 

To extract the light curve, we imaged the field using a 60 s time interval and a mask around the target with the \textsc{casa} \texttt{tclean} task, setting \texttt{niters=10} and \texttt{cycleniter=10}. We selected these values after evaluating random images where the target was either detected or not detected, and ensuring that this choice did not result in false positive detections. Then, we used the \texttt{imfit} task to obtain the peak flux density of the target and the check source, J102348.2+004017, for each 60 s image to investigate possible systematic effects.
To measure the rms noise level around \src\ and check source, we placed four boxes around each source and used the \texttt{imstat} task to average the rms noise level within each box. To build the light curve, we considered a source as detected whenever the flux density value obtained with \texttt{imfit} was $\ge$3$\sigma$; otherwise, we took three times the rms noise level as an upper limit. To take into account possible systematics affecting the FOV, we investigated possible correlations between the light curves of the target and the check source. No evidence of such correlations was found. Finally, we further flagged bad 60 s images and removed intervals with odd results, such as sudden increases in the flux density of J102348.2$+$004017.

\subsection{FAST (Night\,1)}
\label{sec:FAST}
FAST \citep{jiang19,jiang20,qian20} observed \src\ starting on 2021 June 4 at 08:47:45 UTC. The total integration time was 3\,hr (the first two minutes and the last two minutes were dedicated to the injection of an electronic noise signal, referred to as a CAL). The observation was planned so as to cover a range of orbital phases from approximately 0.5 to 1.1, and hence to encompass the epoch of passage of the pulsar at the inferior conjunction of the orbit (corresponding to a phase of 0.75). This strategy should allow us to minimise the impact of eclipses of the radio signal caused by the gas out-flowing from the companion star due to irradiation from the pulsar wind. The observation sampled a total of 21 low mode episodes according to the simultaneous \xmm\ observation (see Fig.\,\ref{fig:timeseries1}).
The central observing frequency was 1.25\,GHz, with a range from 1.05\,GHz to 1.45\,GHz, including a 20\,MHz band edge on each side. The average system temperature was 25\,K. The recorded FAST data stream for pulsar observations is a time series of total power per frequency channel, stored in \texttt{psrfits} format \citep{hotan04} from a ROACH-2-based back-end\footnote{\url{https://casper.astro.berkeley.edu/wiki/ROACH-2_Revision_2}}, which produces 8-bit sampled data over 4k frequency channels at a cadence of 49.152\,$\mu$s. 

We searched for pulsations with either a dispersion signature or instrumental saturation in all data collected during the observation. We performed four different types of data processing.

\textit{Data folding based on the ephemeris derived from the X-ray data}: We analysed the data using the ephemeris derived from the strictly simultaneous X-ray data collected by \xmm\ \citep{miravalzanon22}. We separately folded the FAST data from the entire observation and those obtained by stacking the data acquired during all the periods of high X-ray mode and all the periods of low X-ray mode. However, no radio pulses were detected with S/N$>$6 in either case.

\textit{Dedicated search for persistent periodic radio pulses}: For the sake of robustness, we created de-dispersed time series for each pseudo-pointing over a dispersion measure (DM) range between 3\,pc\,cm$^{-3}$ and 100\,pc\,cm$^{-3}$. This range covers all uncertainties of the semi-blind search, as \src\ was discovered at a DM of 14.3\,pc\,cm$^{-3}$ \citep{Archibald2009}. We set the step size between subsequent trial DMs equal to the intra-channel smearing over the entire band. This ensures that any trial DM deviating from the source DM by $\Delta$DM does not cause extra smearing beyond the intra-channel smearing\footnote{The de-dispersion scheme was generated using the \texttt{DDplan.py} routine in the PulsaR Exploration and Search TOolkit pipeline (\texttt{presto}; \citealt{ransom2001}).}. We performed a `jerk' acceleration search (\citealt{andersen18}; see also \citealt{wang21}) 
for periodic signals in the de-reddened power spectrum of each de-dispersed, barycentred time series. We restricted the search to various frequency ranges around the nominal spin frequency of \src, and allowed the power of the highest harmonic component of the signal to drift in the Fourier domain up to a maximum number of 200 frequency bins and 50 frequency derivative bins.  The powers of the first eight harmonics were summed (in powers-of-two harmonics). All identified candidate periodicities underwent a sifting procedure to reject less significant, duplicated (i.e. detected at different DMs and accelerations), and/or harmonically-related candidates. Using the \texttt{prepfold} tool, the data were folded at the nominal period, period derivative and DM of the sifted candidates. Corresponding diagnostic plots including the integrated pulse profile, the signal intensity as a function of time and frequency, and the DM versus $\chi^2$ values were generated and saved for visual inspection. The candidates were judged based on persistent and broadband emission, as well as distinct peaking of the signal's significance at specific DM values. No signal with DM, period, and period derivative close to the expected values was detected.

\textit{Search for single pulses}: We used the dedicated search scheme in (ii) to de-disperse the time series. We then applied 14 box-car width-match filter grids distributed in logarithmic space from 0.1 to 30\,ms. A zero-DM matched filter was used to mitigate radio frequency interference (RFI) in the blind search. All the resulting candidate plots were visually inspected, but most candidates were determined to be RFI. No pulsed radio emission with a dispersive signature was detected with S/N$>$5.

\textit{Search for baseline saturation}: We acknowledge that FAST would become saturated if the radio flux density reached levels as high as 1 kJy to 1 MJy. Therefore, we also searched for saturation signals in the recorded dataset. We looked for epochs where at least half of the channels satisfied one of the following conditions: (1) the channel was saturated (255 value in 8-bit channels), (2) the channel had a zero value, or (3) the rms deviation of the bandpass was $<$2. We did not find any instances of saturation lasting $>$0.5\,s during the observational campaign.

We then calibrated the noise level of the baseline with noise CAL injection, and derived a 5-$\sigma$ flux density upper limit for persistent radio pulsations of 2.0$\pm$0.5\,$\mu$Jy at 1.25\,GHz considering the entire integration time of 2.8\,hr, assuming a pulse duty cycle between 0.05--0.3 (which is typical for millisecond pulsars; see the ATNF pulsar catalogue; \citealt{manchester05}) and disregarding the possible scintillation effect. Within the uncertainty of the flux measurements and stability threshold, no flux variations were seen during the observation, consistent with an apparent lack of scintillation. The 5-$\sigma$ flux upper limits on the pulsed flux density of \src\ during the high and low modes can be evaluated by scaling the above-mentioned limit for the specific integration time spans during the high and low-mode periods in the X-rays: we obtain upper limits of 2.2$\pm$0.6\,$\mu$Jy in the high mode (2.3\,hr) and 4.7$\pm$1.0\,$\mu$Jy in the low mode (0.5\,hr).

\subsection{NICER/XTI observations (Night\,2)}
The X-ray Timing Instrument (XTI) mounted on NICER \citep{gendreau12} observed \src\ between 2021 June 26 at 21:56:04 UTC and June 27 at 01:33:51 UTC, for an exposure time of $\sim$6.1\,ks. The raw, unfiltered data were processed and analysed using tools within the \textsc{nicerdas} software package and the latest version of the calibration files at the time of the analysis (\textsc{caldb} v. 20221001)\footnote{We followed the standard procedures outlined on the NICER data analysis webpage (\url{https://heasarc.gsfc.nasa.gov/docs/nicer/analysis_threads/}).}. We extracted the background-subtracted light curve in the 0.5--10\,keV energy range with a time bin of 10\,s using the {\sc nicerl3-lc} pipeline. This pipeline utilises the `Space Weather' background model, that is to say, it estimates the background level on the basis of the local spacecraft environment (see Fig.\,\ref{fig:timeseries2}, first panel).

\subsection{Swift/XRT observations (Nights 1 and 2)}
The \swift\ \citep{gehrels04} X-ray Telescope (XRT; \citealt{burrows05}) observed \src\ in the photon-counting mode (time resolution of 2.5\,s) on both Night\,1 and Night\,2. 
The first observation started on 2021 June 3 at 18:29:46 UTC and consisted of 11 snapshots, with a total elapsed time of $\simeq$51.3\,ks and an effective exposure time of $\sim$17.9\,ks. The second observation started on 2021 June 26 at 22:19:35 UTC and consisted of three snapshots, resulting in a total elapsed time of $\simeq$7.3\,ks and an effective exposure time of $\sim$2.4\,ks. The raw, level-1 data were processed and screened using standard criteria. Source photons were collected from a circular aperture of radius 47.2\arc\ centred on \src\ and background photons were collected from an annular region with inner and outer radii of 94.4\arc\ and 188.8\arc, respectively, also centred on \src. 
%\src\ was detected at an average net count rate of 0.23$\pm$0.01 counts\,s$^{-1}$ over the energy range 0.3--10\,keV. %corresponding to a signal-to-noise ratio of S/N$\simeq$19. 
%Given the smaller photon counting statistics with respect to NICER data, the XRT data were used only to disentangle additional time intervals of high and low X-ray modes in the time series that were not covered by the NICER observations. 
We extracted a background-subtracted time series with a time bin of 100\,s, suitable to sample adequately the X-ray mode switching (see Figs.\,\ref{fig:timeseries1} and \ref{fig:timeseries2}).

\subsection{Swift/UVOT observations (Nights 1 and 2)}
The \swift\ Ultra-Violet/Optical Telescope (UVOT; \citealt{roming05}) observed \src\ simultaneously with the XRT. The observations were performed in both image and event modes using the UVM2 filter (central wavelength 2260\,\AA; full width at half maximum 527\,\AA). Data were processed and analysed in the standard way. For the source photons, we used a circular extraction region of radius of 5\arc\ centred on the source position. For the background photons, we adopted a circular extraction region of radius 20\arc\ away from source. For the data in image mode, photometry was carried out using the \texttt{uvotsource} task.
%\src\ was detected at an average magnitude of UVM2 = 16.91$\pm$0.05 (in the Vega system). %(S/N$\simeq$40.5).
For the data in event mode (time resolution of 11\,ms), we used the \texttt{coordinator} tool to convert raw coordinates to detector and sky coordinates, \texttt{uvotscreen} to filter out hot charge-coupled device (CCD) pixels and obtain a cleaned event list, and \texttt{uvotevtlc} to extract a background-subtracted light curve with a time bin of 120\,s. This time bin provided a sufficiently high S/N to detect any changes in the UV brightness at least during the longest periods of low X-ray mode.

\subsection{REM photometric observations (Night\,2)}
We observed \src\ on Night\,2 with the 60-cm REM telescope \citep{Zerbi2001,Covino2004} located at La Silla Observatory (Chile). We obtained 48 strictly simultaneous exposures with the ROSS2 imager, each with an integration time of 60\,s, with Sloan Digital Sky Survey (SDSS) $griz$ optical filters (Table\,\ref{tab:log}). We reduced the images following standard procedures (bias subtraction and division by a normalised flat-field), and performed aperture photometry with the {\tt PHOT} tool in {\tt IRAF} using an aperture size of 6 pixels. We performed flux calibration using stars in the field from the American Association of Variable Star Observers (AAVSO) Photometric All-Sky Survey (APASS) catalogue\footnote{\url{http://www.aavso.org/download-apass-data}} \citep{Henden2019}. 
We obtained simultaneous $K$-band observations with the REM infrared (REMIR) camera. A total of 300 15 s integration images were acquired, which we combined 5 by 5 to correct for the background contribution. Aperture photometry was performed with {\tt IRAF} as described above. The images were flux-calibrated against a selection of stars in the field from the 2MASS point source catalogue\footnote{\url{https://irsa.ipac.caltech.edu/applications/2MASS/IM/}}. 

\subsection{VLT/FORS2 polarimetric observations (Night\,2)}\label{sec:pol}
We observed \src\ with the FOcal Reducer/low dispersion Spectrograph 2 (FORS2; \citealt{appenzeller98}) mounted on the 8.2-m VLT at Paranal Observatory (Chile) in polarimetric mode using the optical filter $R\_SPECIAL +76$ ($R$; central wavelength 655\,nm, full width at half maximum 165\,nm). The observations were performed on Night\,2 under thin cirrus clouds conditions, with seeing varying from $0.4$\arc\ to $0.9$\arc. 
A total of 156 short (10\,s) exposures were taken, resulting in an observing time of $\sim$2.41\,hr, overheads included (about half the orbital period). 

A Wollaston prism was inserted in the optical light path to split
the incident radiation into two orthogonally polarised light beams called ordinary ($o$) and extraordinary ($e$) beams. A Wollaston mask was introduced to avoid overlap of the beams on the CCD. In addition, a rotating half-wave plate (HWP) was inserted, allowing images to be taken at four different angles ($\Phi_i$) of the plate with respect to the telescope axis: $\Phi_{i}=22.5^{\circ}(i-1), \, i=1, 2, 3, 4$. This step is essential to obtain a polarisation measurement, since the set of images at the four different angles must be combined to evaluate the level of LP (as described below). Therefore, a total of 39 polarisation epochs were acquired with the programme. All images were processed by subtracting an average bias frame and dividing by a normalised flat field. Aperture photometry was performed using the {\tt daophot} tool \citep{Stetson1987}, using an aperture of 6 pixels. 
The normalised Stokes parameters $Q$ and $U$ for the LP were evaluated following the methods described by \cite{baglio20} and references therein.
These values were then corrected for instrumental polarisation by observing a non-polarised standard star\footnote{\url{ https://www.eso.org/sci/facilities/paranal/instruments/fors/tools/FORS1_Std.html}} (WD 1620-391) using the same setup on the same night. A polarisation level of $<0.2\%$ was measured, which is consistent with the typical instrumental polarisation level expected for the FORS2 instrument as reported on the ESO website\footnote{\url{http://www.eso.org/observing/dfo/quality/FORS2/reports/FULL/trend_report_PMOS_inst_pol_FULL.html}} ($<$0.3\% in all optical bands). The Stokes parameters evaluated for the non-polarised standard star were then subtracted from the Stokes parameters of \src\ at each epoch.

We then used an algorithm to estimate the level and angle of LP. This algorithm is based on the evaluation of the quantity $S(\Phi)$ for each of the HWP angles (e.g. \citealt{Serego} and references therein; see also \citealt{Covino1999,baglio20}). $S(\Phi)$ is the component of the Stokes vector that describes the LP along the $\Phi$ direction (e.g. $S(0^{\circ})=Q$). It is defined as
%%%%%%%%%%%%%%%%%%%%%%%%%%%%%%%%%%%%%%%%%%%%%%
\begin{equation}\label{eq_S}
S(\Phi)=\left( \frac{f^{o}(\Phi)/f^e(\Phi)}{f^o_u(\Phi)/f^e_u(\Phi)}-1\right)\,\left( \frac{f^{o}(\Phi)/f^e(\Phi)}{f^o_u(\Phi)/f^e_u(\Phi)}+1\right)^{-1},
\end{equation}
%%%%%%%%%%%%%%%%%%%%%%%%%%%%%%%%%%%%%%%%%%%%%%
where $\Phi$ is the angle of the HWP, $f^{o}(\Phi)$ and $f^{e}(\Phi)$ are the ordinary and extraordinary fluxes for a given angle $\Phi$, and $f^o_u(\Phi)$ and $f^e_u(\Phi)$ are the same quantities evaluated for the non-polarised standard star. $S(\Phi)$ is related to the polarisation level $P$ and angle $\theta$ by $S(\Phi) = P\, \cos 2(\theta-\Phi)$.
The parameters $P$ and $\theta$ were estimated by maximising the Gaussian likelihood function and using this as the starting point for a Markov chain Monte Carlo (MCMC) procedure \citep{Hogg&Foreman2018} (see \citealt{baglio20} for further details on the algorithm). The best fit parameters were then calculated as the median of the marginal posterior distributions of these parameters, while the uncertainties were estimated as the 16--84th percentiles of the distributions. For non-detections, the 99.97\% percentile of the posterior distribution of the parameter $P$ was used to calculate an upper limit.  

After fitting the $S(\Phi)$ curve, we obtained an LP value that is already corrected for instrumental effects. However, the polarisation angle still needs to be corrected using the Stokes parameters of a polarised standard star. For this purpose, we observed the star Hiltner 652 and measured its Stokes parameters to determine its polarisation angle. This was then compared to the tabulated value, and a difference of $1.71^{\circ}\pm 0.30^{\circ}$ was applied to the polarisation angle of \src. However, the resulting LP will still include an interstellar component.
Evaluating the interstellar contribution is not straightforward. In principle, we could use a group of reference stars in the same field as the target in Eq.\,\ref{eq_S}, under the assumption that these stars are non-polarised. This way, the LP of the target would be corrected for both instrumental and interstellar contributions. However, it is important to ensure that the chosen reference stars all have the same level of polarisation, which would be a good indication of their intrinsic non-polarised nature (see e.g. \citealt{baglio20}).
Unfortunately, not only is the FOV of \src\ sparse in the optical, but the very short exposure times (10\,s) exclude all faint ($R>19$) stars from this study. We are therefore left with only one possible comparison star in the field, which makes it difficult to verify whether the chosen star is indeed intrinsically non-polarised.
As a result, we used the fluxes of the non-polarised standard star in Eq.\,\ref{eq_S}. The field star will be used as a comparison for the analysis.

We can evaluate the maximum interstellar contribution to the LP of a source, referred to as $P_{\rm int}$, through considerations of its absorption coefficient in the $V$-band, $A_{\rm V}$: $P_{\rm int}(\%)<3A_{V}$ \citep{Serkowski75}. For \src, the best-fitting absorption column density, $N_{\rm H}$, according to \xmm\ observations is $5.0^{+0.3}_{-0.2}\times 10^{20}$\,cm$^{-2}$ \citep{campana19}. Using the relation between $N_{\rm H}$ and $A_{\rm V}$ in our Galaxy from \cite{Foight2016}, we calculate $A_{\rm V}=0.17\pm0.01$\,mag and therefore $P_{\rm int}<0.52\%$. This is very low and comparable to the instrumental polarisation measured for the instrument configuration. It is also comparable to the upper limit of the interstellar polarisation measured for \src\ \citep{baglio16}.

\subsection{ALMA observations (Night\,2)}
The ALMA observations were conducted using the 12~m array and 38 antennas in the C43-7 configuration (baselines 22--2492\,m), with a FOV of $59.72''$ and a maximum recoverable scale of $11.5''$. The total integration time on source was $\sim$2\,hr and the phase centre was R.A. = 10$^\mathrm{h}$23$^\mathrm{m}$47$^\mathrm{s}$68; Dec. = +00$^{\circ}$38$^{\prime}$41.$^{\prime\prime}$00 (ICRS).
The weather conditions were good and stable, with an average precipitable water vapour of $\sim$1.2\,mm and an average system temperature of $\simeq$65\,K. The band-pass and flux calibrators used were J1058$+$0133 and J1256$-$0547, while J1028$-$0236 was used as the phase calibrator. 
The continuum image was obtained by averaging the four base-bands of width 1.875\,GHz, each centred at 90.5003\,GHz, 92.5003\,GHz, 102.5003\,GHz, and 104.5003\,GHz, covering a total bandwidth of 7.5\,GHz.  
No spectral lines were detected within the observed frequency range above the 3$\sigma$ noise level per spectral channel of $\sim$0.48\,\mJy.

Data reduction was carried out following standard procedures using the ALMA pipeline within \textsc{casa} (v. 6.1.1-15)\footnote{\url{https://casa.nrao.edu}}.  
Continuum images were produced using \texttt{tclean} interactively and applying a manually selected mask. These images were corrected for primary beam attenuation and were produced using both natural weighting and a Briggs \texttt{robust=0} weighting \citep{briggs95}. The former maximises the sensitivity whereas the latter provides higher angular resolution. The mean frequency of the continuum image is 97.51\,GHz, which corresponds to a wavelength of 3.1\,mm. Table\,\ref{tab:alma_obs} lists the synthesised beam and the rms noise of each image.

%%%%%%%%%%%%%%%%%%%%%%%%%%%%%%%%%%%%%%%%%%%%%%
\begin{table}[!t]
\begin{footnotesize}
    \centering
    \caption{ALMA continuum maps.}
    \begin{tabular}{lccc}
    \hline\hline
     Weighting &Synthesised beam & P.A &rms$^{\rm a}$ \\
             &$(''\times'')$  &(\grau) &($\mu$\Jy) \\ 
    \hline
       Natural &$0.87\times0.53$ &$-$74.5   &6.1  \\
       Briggs, \texttt{robust=0}  &$0.55\times0.30$ &$-87.3$ &10.6  \\
         \hline
    \end{tabular}
    \parbox{0.5\textwidth}{
    {\bf Notes.}
    $^{\rm a}$ rms noise level at the phase centre.
    }
     \label{tab:alma_obs}
\end{footnotesize}
\end{table}
%%%%%%%%%%%%%%%%%%%%%%%%%%%%%%%%%%%%%%%%%%%%%%

\section{Results}
\label{sec:results}

\subsection{X-ray and UV emissions}
\label{sec:XUV}
\begin{figure*}
\begin{center}
\includegraphics[width=1.0\textwidth]{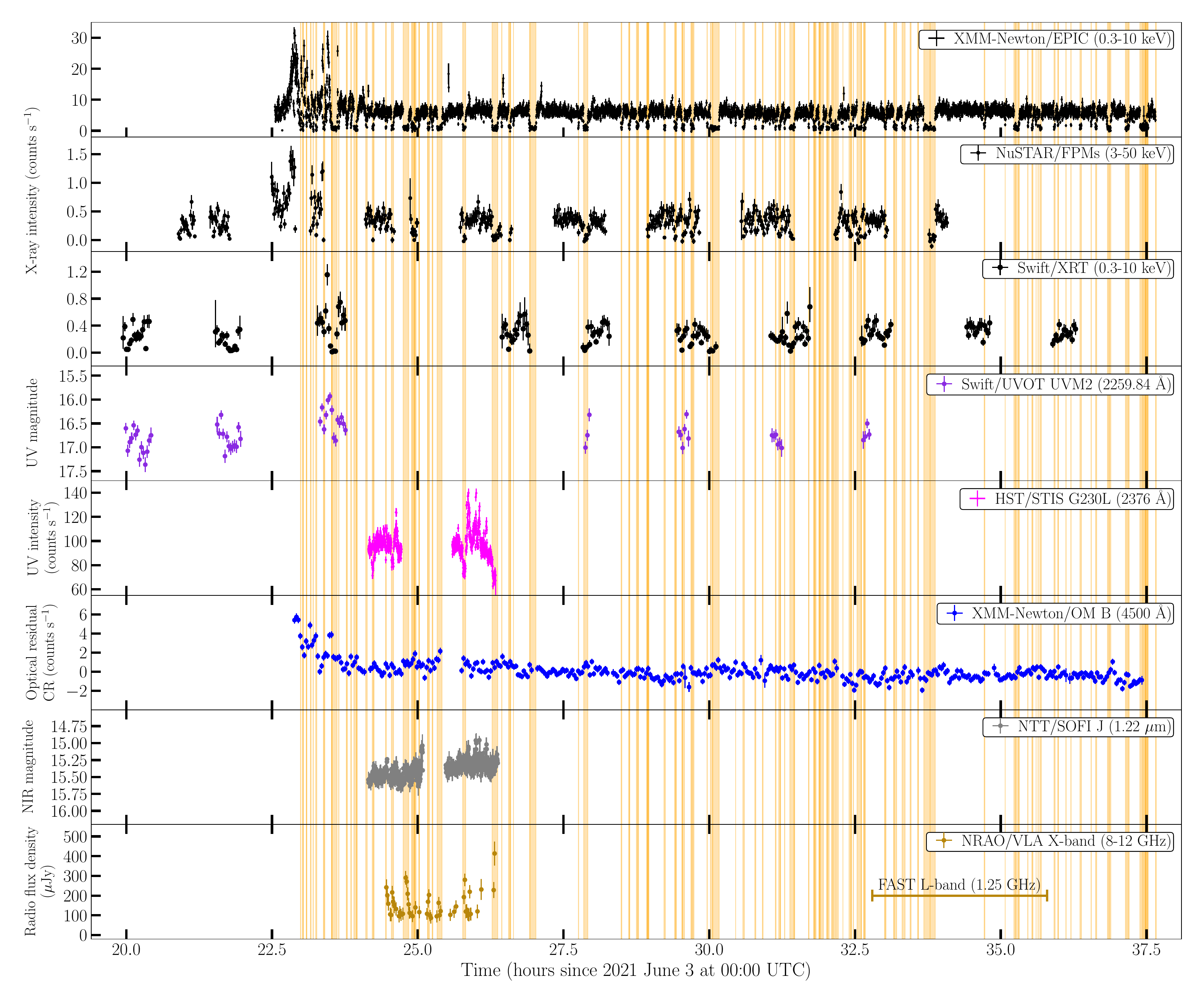}
\vspace{-0.5cm}
\caption{Temporal evolution of the X-ray, UV, optical, NIR, and radio emissions of \src\ on the night of 2021 June 3--4. 
The light curves are shown in decreasing order of energy band from top to bottom, with the \xmm/EPIC one as the reference for all others (\nustar, HST/STIS, \xmm/OM, NTT/SOFI, and VLA).
The \xmm\ X-ray light curve is shown over the time interval in which the three EPIC instruments collected data simultaneously.
The OM residual count rates were obtained after removing the orbital modulation from the time series (see the main text for details).
For displaying purposes, only VLA detections are shown. The dark yellow segment in the bottom panel indicates the time range covered by the FAST observation. The orange shaded areas mark the time intervals of the low X-ray mode detected by \xmm/EPIC covered by observations with other instruments.} 
\label{fig:timeseries1}
\end{center}
\end{figure*}

The X-ray light curve extracted from \xmm\ data on the first night reveals flaring activity over a time span of $\sim$1.5\,hr at the beginning of the observation. A total of 288 switches between high and low modes were observed during the observation (see Fig.\,\ref{fig:timeseries1}). Based on our thresholds for the EPIC net count rates in the different modes (see \citealt{miravalzanon22}), we estimated that \src\ spent $\simeq$65, 17 and 1\% of the time in the high, low and flaring modes, respectively, and the rest of the time switching between modes. 
The good time intervals corresponding to the high and low modes were then used to extract the background-subtracted spectra, response matrices, and ancillary files separately for each mode.
The \nustar\ light curve is characterised by a lower S/N than the \xmm/EPIC light curve. Nevertheless, it is possible to recognise the same flaring episodes and mode switchings detected using \xmm\ (see Fig.\,\ref{fig:timeseries1}).
We used the good time intervals for the high and low modes selected by the strictly simultaneous \xmm/EPIC light curves to extract the background-subtracted spectra, response matrices, and ancillary files separately for each mode.

\begin{figure*}
\begin{center}
\includegraphics[width=1.0\textwidth]{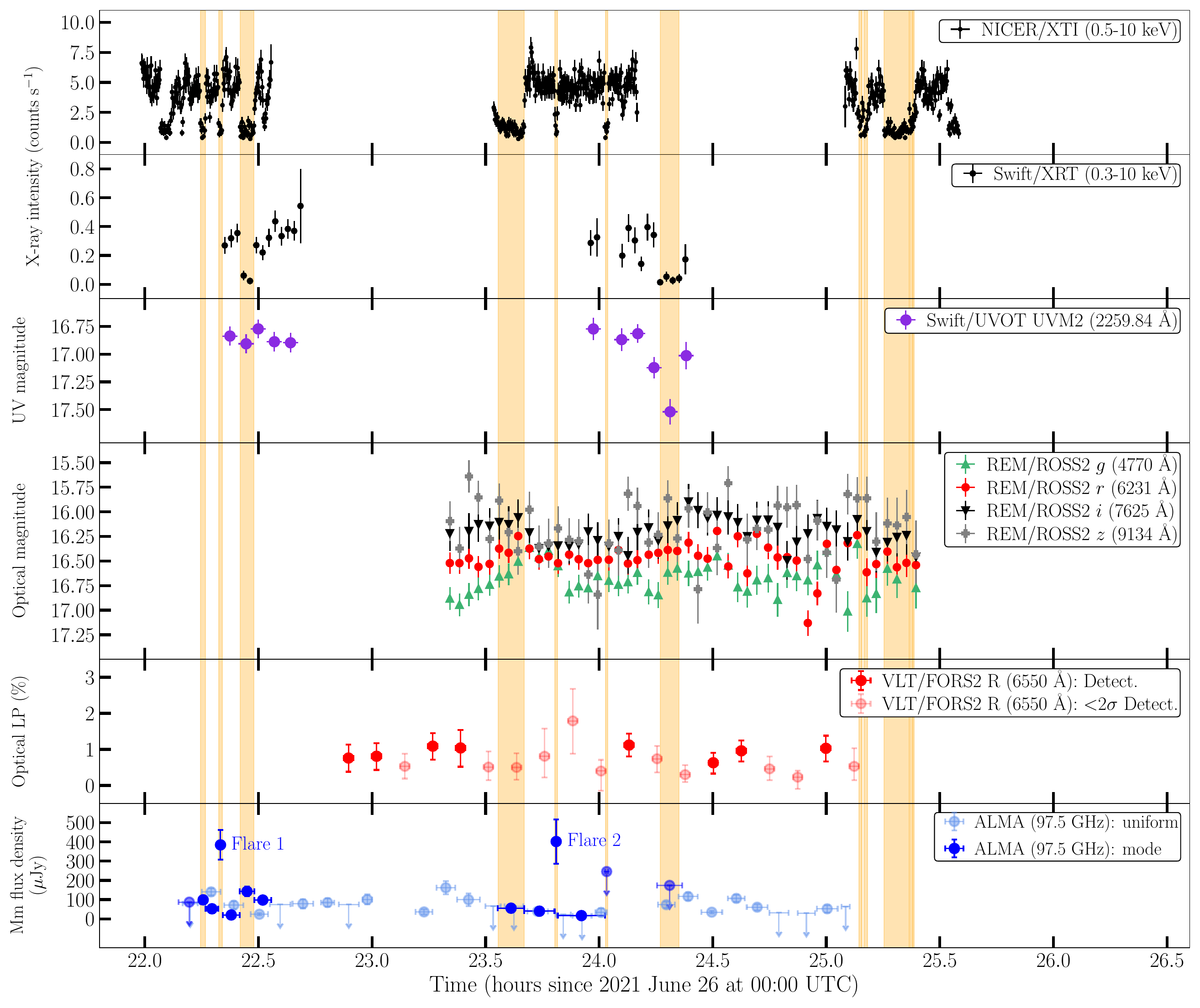}
\caption{Temporal evolution of the X-ray, UV, optical, and millimetre emissions and of the optical LP of \src\ on the night of 2021 June 26--27. The light curves are shown in decreasing order of energy band from top to bottom (NICER, \swift/XRT, \swift/UVOT, REM, VLT/FORS2 polarisation, and ALMA). In the bottom panel, the ALMA time series are plotted using either bins of length 4-5\,min (cyan) or bins of variable length corresponding to the duration of the X-ray mode episodes identified by the strictly simultaneous NICER and \swift\ data (blue). In some panels, the sizes of the markers are bigger than the uncertainties and/or the time bins. Arrows indicate 3$\sigma$ upper limits. The orange shaded areas mark the time intervals of low X-ray mode detected by NICER and \swift/XRT covered by observations with other instruments. A plot of the X-ray and millimetre light curves around the epoch of the millimetre flares is shown Fig.\,\ref{fig:timeseries_alma_flares}.} 
\label{fig:timeseries2}
\end{center}
\end{figure*}

The X-ray light curve extracted from NICER data on the second night, shown in the top panel of Fig.\,\ref{fig:timeseries2}, reveals 22 instances of X-ray mode switching, without any obvious flaring episodes. We applied intensity thresholds to the light curve to select the time intervals corresponding to the high ($>$2.5\,counts\,s$^{-1}$) and low ($<$2.5\,counts\,s$^{-1}$) X-ray modes. Based on this selection, we estimate that \src\ spent $\sim$70\% of the time in the high mode and $\sim$30\%  of the time in the low mode. We also detected two periods of X-ray low mode in the \swift/XRT data, one of which was also covered by the NICER observations (see the second panel of Fig.\,\ref{fig:timeseries2}).

The UV light curves exhibit substantial variability. Specifically, the HST light curve on Night\,1 displays flaring activity as well as dips in the count rate that coincide with the low-mode episodes observed in X-rays (Fig.\,\ref{fig:timeseries1}; see also \citealt{miravalzanon22,jaodand21}). 
A dip in the UV brightness by $\sim$0.7\,mag is also clearly visible in the \swift/UVOT data during the third snapshot of Night\,2. The decrease occurs over a span of several minutes during an episode of low mode (Fig.\,\ref{fig:timeseries2}).

%%%%%%%%%%%%%%%%%%%%%%%%%%%%%%%%%%%%%%%%%%%%%%%%%%%%%
%%%%%%%%%%%%%%%%%%%%%%%%%%%%%%%%%%%%%%%%%%%%%%%%%%%%%

%%%%%%%%%%%%%%%%%%%%%%%%%%%%%%%%%%%%%%%%%%%%%%%%%%%%%
\begin{figure*}
\begin{center}
\includegraphics[width=1.0\textwidth]{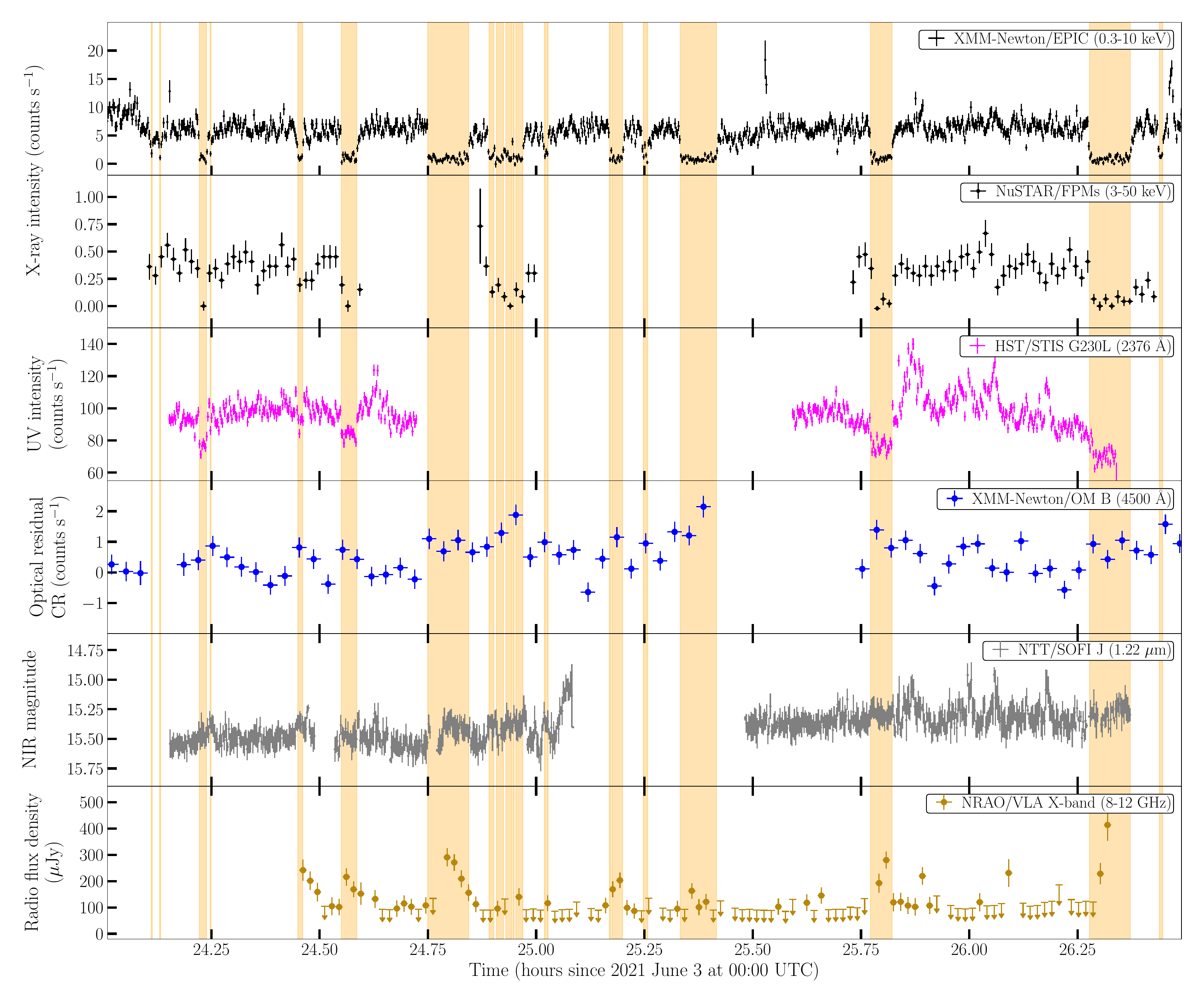}
\vspace{-0.5cm}
\caption{Zoom-in of the time series collected on Night\,1 over the time span of maximum overlap among the different telescopes. Arrows indicate 3$\sigma$ upper limits.} 
\label{fig:timeseries1_zoom}
\end{center}
\end{figure*}
%%%%%%%%%%%%%%%%%%%%%%%%%%%%%%%%%%%%%%%%%%%%%%%%%%%%%

%%%%%%%%%%%%%%%%%%%%%%%%%%%%%%%%%%%%%%%%%%%%%%%%%%%%%
%%%%%%%%%%%%%%%%%%%%%%%%%%%%%%%%%%%%%%%%%%%%%%%%%%%%%

%%%%%%%%%%%%%%%%%%%%%%%%%%%%%%%%%%%%%%%%%%%%%%%%%%%%%
\begin{figure}
\begin{center}
\vspace{-1cm}
\includegraphics[width=0.52\textwidth]{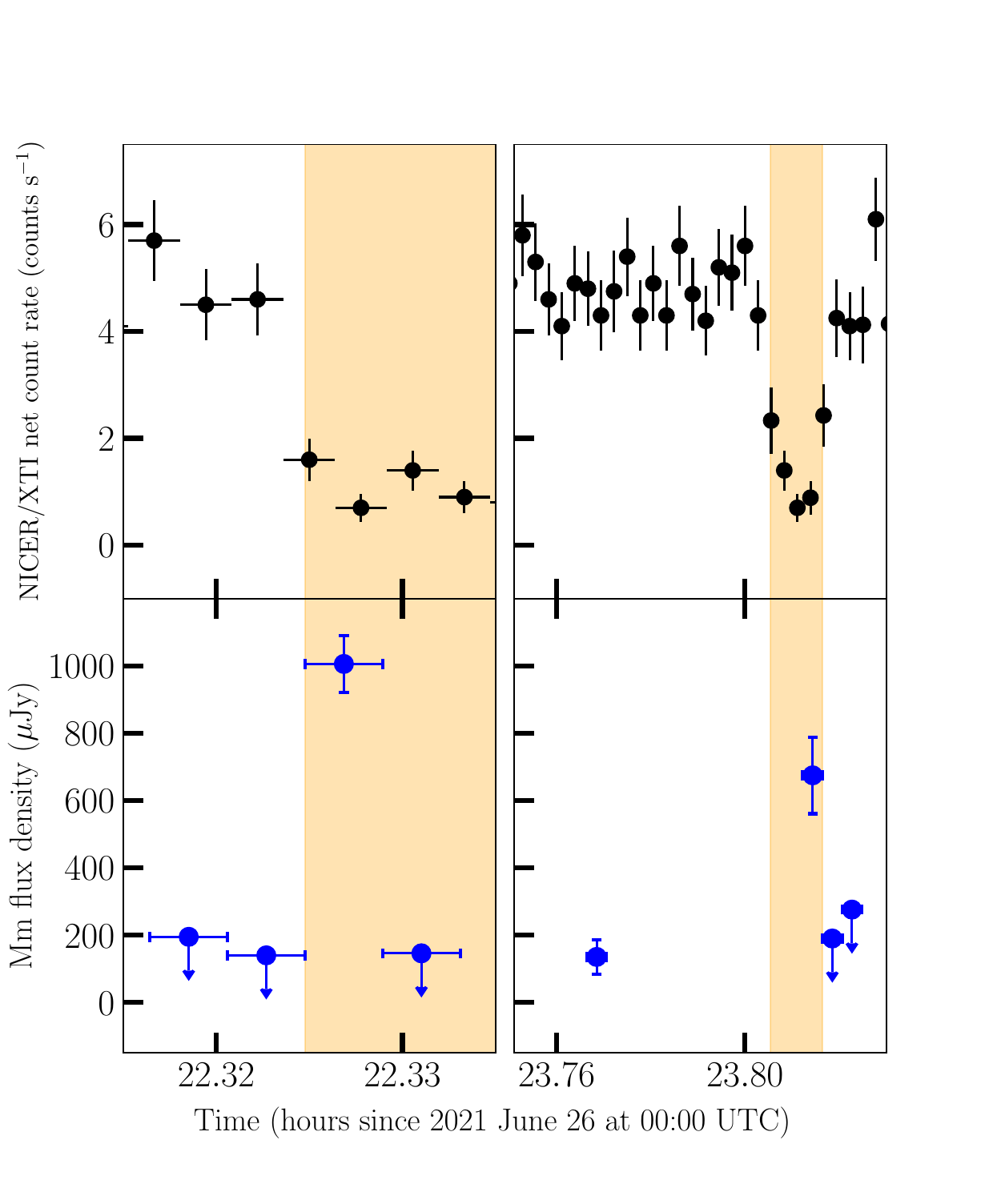}
\vspace{-1cm}
\caption{X-ray (top) and millimetre (bottom) time series collected around the epochs when the two millimetre flares were detected. The X-ray light curve is binned at a time resolution of 10\,s, and the millimetre light curve is binned at 15\,s. Arrows indicate 3$\sigma$ upper limits. Note the lack of coverage in the millimetre band for Flare 2 before and during the switch from the high to the low mode in the time interval between two consecutive ALMA scans. Note also the different scales on the horizontal axis in the panels on the left and on the right.} 
\label{fig:timeseries_alma_flares}
\end{center}
\end{figure}
%%%%%%%%%%%%%%%%%%%%%%%%%%%%%%%%%%%%%%%%%%%%%%%%%%%%%

\subsection{Optical/NIR photometric properties}
On the first night, the source is brighter at the beginning than in the rest of the observation (reaching $B\simeq17-16.5$\,mag), displaying a smooth quasi-sinusoidal modulation at the orbital period with a semi-amplitude of $\simeq$0.2\,mag around an average value of $B\simeq17.3$ mag. 
We fit a sinusoidal function to the light curve by fixing the period and reference phase of the sinusoid to the binary orbital period and the phase value obtained from the time of passage of the pulsar through the ascending node, respectively, as derived by \cite{miravalzanon22}. Then, we subtracted the best-fitting function from the observed count rates.
The resulting detrended light curve is shown in Fig.\,\ref{fig:timeseries1}. After initial flaring activity resembling the X-ray behaviour, the optical intensity appears to be flickering, but without any sign of mode switching.

Figure\,\ref{fig:timeseries1} also shows the NIR light curve. Variability in the form of dips and flickering is evident. The fractional rms magnitude deviation of the light curve, computed following the method described by \cite{vaughan03}, is (10.1$\pm$0.2)\%.
No NIR flare was observed at the switch from the low to the high X-ray mode, as previously reported \citep{papitto19,baglio19}. Furthermore, we did not detect any signs of mode switching in the light curve. The average magnitudes measured separately during the high and low modes are compatible within 1$\sigma$ ($J=15.39\pm 0.16$\,mag and $J=15.37\pm 0.13$\,mag in the high and low modes, respectively), indicating that there is no significant correlation between the NIR variability and mode switching.

On Night\,2, the optical light curves show clear variability on short timescales (Fig.\,\ref{fig:timeseries2}, fourth panel). 
Unlike what was seen on Night\,1, no modulation at the system orbital period is detected.
The average magnitudes are $g=16.65\pm0.02$ mag, $r=16.45\pm0.02$ mag, $i=16.20\pm 0.03$ mag, $z=16.09\pm0.03$ mag (in the AB photometric system), all marginally consistent with those reported by \cite{cotizelati14, baglio16}. 
To estimate the amount of intrinsic variability, we evaluated the fractional rms magnitude deviation of the light curves following \cite{vaughan94}. We find a low-significance fractional rms of $6.2\pm3.4\%$, $6.9\pm2.3\%$ and $12.7\pm4.7\%$ at a time resolution of 160\,s in the $g$, $r$ and $z$ bands, respectively, while no excess intrinsic variability could be detected in the $i$ band. 
This fractional rms is comparable to what has been observed for a few black hole XRBs in the optical band on a similar timescale, such as GX\,339-4 \citep{Gandhi2009,Gandhi2010} and MAXI\,J1535$-$571 \citep{Baglio2018}. In these systems, the variability has been attributed to the emission of a flickering jet that contributes up to optical frequencies. In contrast, when accretion dominates, lower values of fractional rms are typically measured, as seen in the case of the black hole XRB GRS\,1716$-$249 \citep{Saikia22}. 

Unfortunately, most of the $K$-band images are of bad quality, resulting in an empty FOV.  We stacked together three of the combined images in which the field was visible and \src\ was detected (so a total of 15$\times$15\,s integration images, all taken during the high mode), and performed the calibration against a selection of stars in the field from the      2MASS point source catalogue. We obtained an average magnitude of $K=14.41 \pm 0.29$\,mag, which is consistent with previous values reported in the literature (see e.g. \citealt{cotizelati14,baglio16}).

\subsection{Optical polarimetric properties}\label{res: polarimetry}
Figure\,\ref{fig:timeseries2} shows the evolution of the optical LP level on the second night of observations. To improve the S/N of the observations without losing too much information on the short-term variability of the LP, we combined the measurements so that each LP data point covers 80\,s (four HWP angles $\times$ 10\,s of integration $\times$ two sets). 
We define an LP measurement as a detection when it reaches a confidence level of greater than 2$\sigma$. Using this threshold, we obtain a total of 8 LP detections and 11 upper limits at a 99.97\% confidence level (see Table\,\ref{tab:polarisation}). The average level of LP, calculated as the average of all 2$\sigma$-significant detections, is $0.92\pm0.39\%$. This is consistent with previous results in the same band \citep[$R$-band;][]{baglio16}. 
The polarisation angle is well constrained at a value of $(6^{+18}_{-15}$)$^{\circ}$ (corrected for the polarisation angle of the standard star).

%%%%%%%%%%%%%%%%%%%%%%%%%%%%%%%%%%%%%%%%%%%%%%%%%%%%%
\begin{table}
    \centering
    \footnotesize
    \caption{LP detections and angles with a significance $>$2$\sigma$.}
    \begin{tabular}{lccc}
    \hline\hline
    Epoch (MJD)&LP ($\%$) &PA ($^{\circ}$) & Upper limit ($\%$) \\[5pt]
    \hline
    59391.95402 (?) & $0.76^{+0.38}_{-0.37}$ & $27.32^{+15.41}_{-14.08}$ & \\[5pt]
    59391.95917 (?)& $0.81^{+0.38}_{-0.36}$ & $-3.57^{+13.33}_{-12.62}$ & \\[5pt]
    59391.96436 (?) &      --                &           --              & 1.77 \\[5pt]
    59391.96949 (?) & $1.09^{+0.37}_{-0.36}$ & $-2.92^{+9.28}_{-9.37}$ & \\[5pt]
    59391.97459 (?)& $1.04^{+0.52}_{-0.50}$ & $1.56^{+13.82}_{-14.64}$ & \\[5pt]
    59391.97971 (?) &      --                &           --              & 1.99 \\[5pt]
    59391.98485 (L) &      --                &           --              & 1.82 \\[5pt]
    59391.99000 (H)&      --                &           --              & 3.42 \\[5pt]
    59391.99517 (H)&      --                &           --              & 4.77 \\[5pt]
    59392.00034 (H)&      --                &           --              & 2.33 \\[5pt]
    59392.00549 (H) & $1.12\pm 0.32$ & $21.96^{+8.14}_{-8.41}$ & \\[5pt]
    59392.01063 (?)&      --                &           --              & 2.02\\[5pt]
    59392.01578 (?) &      --                &           --              & 1.22 \\[5pt]
    59392.02094 (?) & $0.63^{+0.30}_{-0.28}$ & $-6.46^{+12.98}_{-12.57}$ & \\[5pt]
    59392.02608  (?)& $0.96\pm 0.29$ & $1.38^{+7.58}_{-7.56}$ & \\
    59392.03126 (?)&      --                &           --              & 1.61\\[5pt]
    59392.03643 (?) &      --                &           --              & 1.27\\[5pt]
    59392.04160  (?)& $1.03^{+0.36}_{-0.35}$ & $9.39^{+10.11}_{-10.82}$ &\\[5pt]
    59392.04680 (H)&      --                &           --              & 2.25 \\[5pt]
      \hline
    \end{tabular}
    \parbox{0.5\textwidth}{
    {\bf Notes.}
    For non-detections, upper limits are quoted at a confidence level of 99.97\%. The first column indicates, in parentheses, which mode the MJD refers to (`L' for low mode, `H' for high mode, `?' when unknown due to the lack of simultaneous X-ray observations).
    }
     \label{tab:polarisation}
\end{table}
%%%%%%%%%%%%%%%%%%%%%%%%%%%%%%%%%%%%%%%%%%%%%%%%%%%%%

When detected, the LP appears fairly constant throughout the observations.
We investigated any potential changes in the LP level as a function of the X-ray mode (low/high) by averaging all images taken during the high and low modes. We found that the LP was $0.65\pm0.43$\% (with a 99.97\% confidence level upper limit of $2.12\%$) in the high mode and $0.36^{+0.27}_{-0.35}\%$ (with a 99.97\% confidence level upper limit of $1.58\%$) in the low mode. 

\begin{figure}
    \centering
    \includegraphics[width=0.48\textwidth]{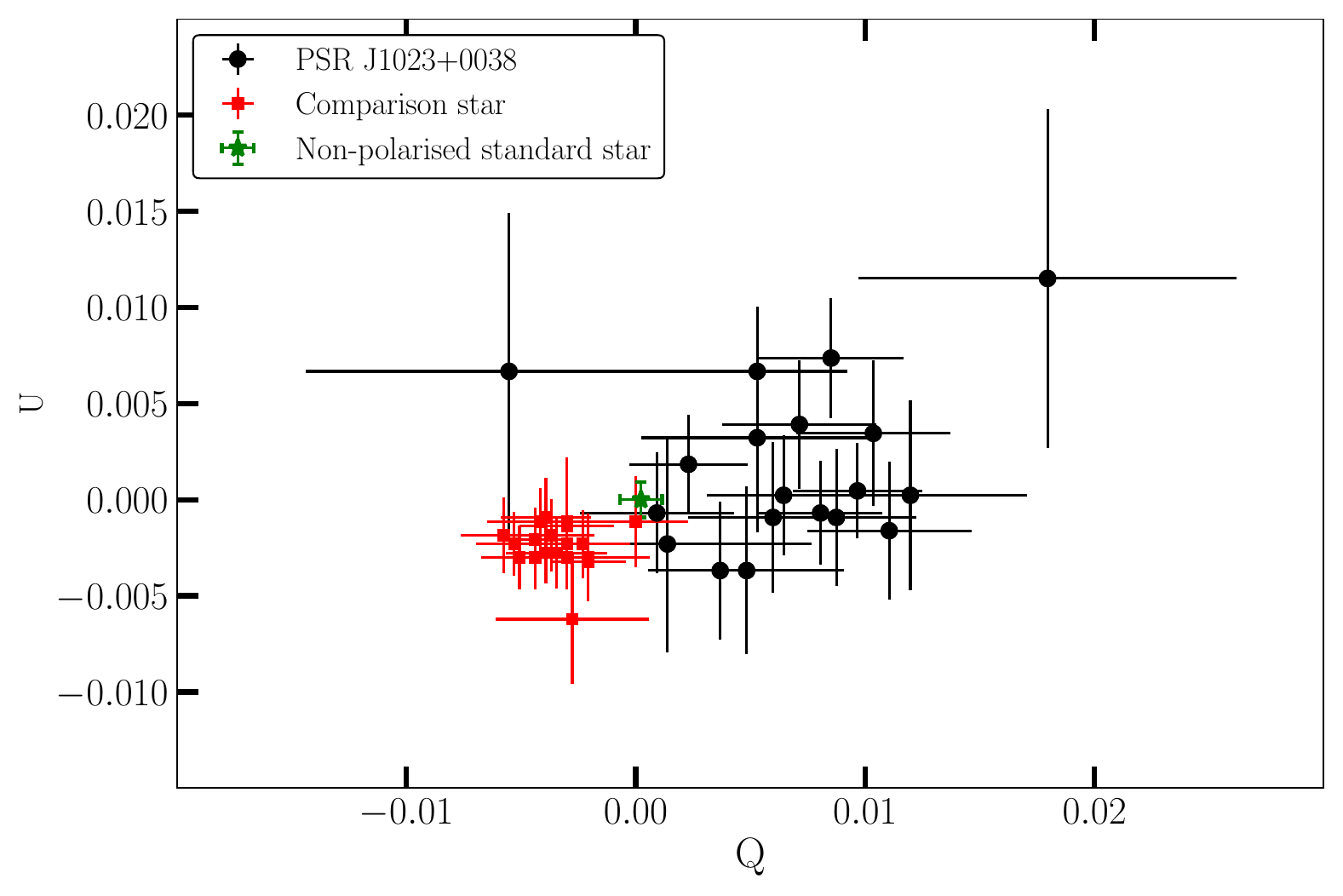}
    \caption{Stokes $Q$ and $U$ measurements for the whole VLT/FORS2 dataset. Black dots, red squares, and the green star indicate the Stokes parameters of \src, a comparison star in the field, and the non-polarised standard star, respectively. }
    \label{fig:Q_U}
\end{figure}

The measured levels of LP are not corrected for interstellar contribution, which is expected to be low according to Serkowski's law ($<$0.52\%; \citealt{Serkowski75}).
Figure\,\ref{fig:Q_U} shows the $Q$ and $U$ Stokes parameters on the $Q-U$ plane for the whole dataset, including \src\ (black dots), a comparison star in the field (red squares), and the non-polarised standard star (green star). The non-polarised standard star is located very close to the origin of the $Q-U$ plane, indicating that the instrumental polarisation of VLT/FORS2 is low, as expected. The Stokes parameters of the comparison star, on the other hand, are scattered around a common point, consistent with the position of the non-polarised standard within $\sim$1.5$\delta$, where $\delta$ is the standard deviation of the points. Any deviation from the position of the non-polarised standard on the plane is a good estimate of the interstellar polarisation of the field (assuming that the comparison star is intrinsically non-polarised).
Interestingly, the Stokes parameters of \src\ are clustered around a common point in a different region of the $Q-U$ plane, with a higher scatter than the comparison star due to larger uncertainties. For the two time series of the $Q$ and $U$ Stokes parameters, we are unable to claim any detection of short-timescale variability.

%%%%%%%%%%%%%%%%%%%%%%%%%%%%%%%%%%%%%%%%%%%%%
%%%%%%%%%%%%%%%%%%%%%%%%%%%%%%%%%%%%%%%%%%%%%

\subsection{Millimetre emission}
\label{sec:mm_results}
\begin{figure*}
    \centering
    \includegraphics[width=1\textwidth]{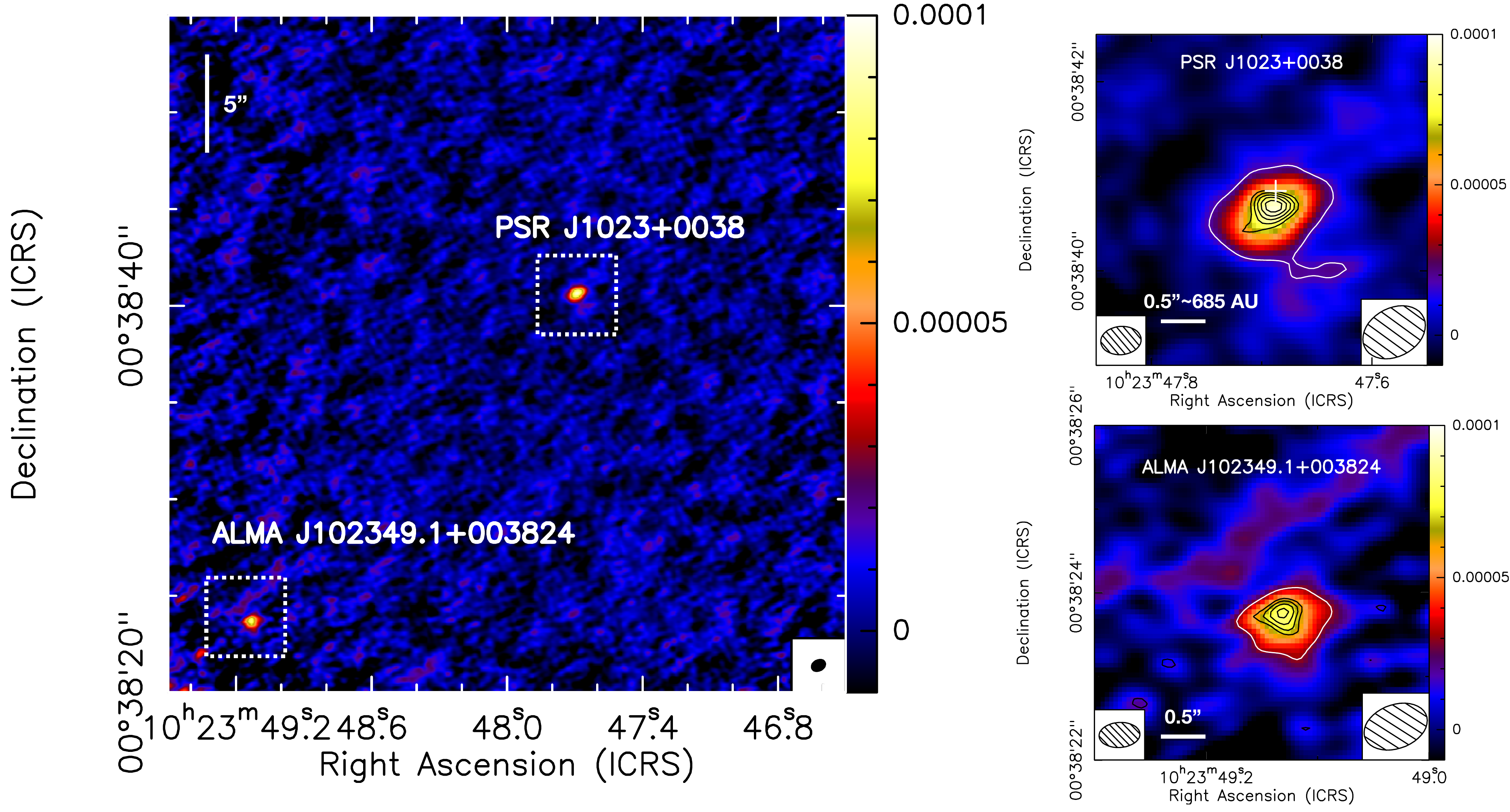}
    \vspace{-0.3cm}
    \caption{ALMA images of the field around \src. \emph{Left:} ALMA 3.1~mm continuum image showing the two detected sources, \src\ and \srcmm. The white dotted boxes display the FOV shown in the right panels. The ALMA synthesised beam ($0.87''\times0.53''$, P.A.=$-74.5$\deg) is shown at the bottom-right corner.
    \emph{Right:} Close-up images of \src\ (top) and \srcmm\ (bottom). In both panels, the colour scale corresponds to the image with natural weighting, with the 3$\sigma$ contour indicated in white ($\sigma$ values are listed in Table\,\ref{tab:alma_obs}. The black contours are derived from the ALMA image obtained with \texttt{robust=0}. Contours range from 3$\sigma$ to 9$\sigma$ in steps of 1$\sigma$, where $\sigma=9.1~\mu$\Jy\ for \src\ and $\sigma=15~\mu$\Jy\ for \srcmm. The white cross depicts the radio position of \src\ \citep{deller12}.
    A faint emission tail is detected at the $3\sigma$ level extending south-west of \src. The synthesised beams for the \texttt{robust=0} and natural images are displayed in the bottom-left and bottom-right corners, respectively (see Table\,\ref{tab:alma_obs}). In all panels, the colour scale is in units of \Jy.}
    \label{fig:alma}
\end{figure*}

Figure\,\ref{fig:alma} shows the ALMA 3.1 mm continuum emission towards \src. 
Two sources were detected within the FOV of ALMA. \src\ was detected at a significance level of $\sim$15$\sigma$. The other source, which we designate \srcmm\ following the naming convention of new sources discovered by ALMA, is located $\sim$22\arc\ south-east of \src\ and was detected at a significance level of $\sim$9$\sigma$. The panels on the right of the figure display the zoom-in maps around each source, with the continuum image obtained with natural weighting displayed in colour scale and the black contours derived from the image with \texttt{robust=0}. Table\,\ref{tab:alma_res} reports the peak coordinates, the peak intensity, the integrated flux, and the deconvolved size obtained by fitting a 2D Gaussian function to the high-resolution image (\texttt{robust=0}).

%%%%%%%%%%%%%%%%%%%%%%%%%%%%%%%%%%%%%%%%%%%%%%
\begin{table*}[!h]
\begin{footnotesize}
    \centering
    \caption{Parameters of the sources detected at 3.1~mm with ALMA.}
    \begin{tabular}{lcccccc}
    \hline\hline
     Source &R.A (ICRS) &Dec. (ICRS) &$I_{\nu}$$^{\rm a}$ &$S_{\nu}$$^{\rm b}$ &Deconvolved Size$^{\rm c}$ &P.A \\
     &(hh:mm:ss) &(\deg:$'$:'') &($\mu$\Jy) &($\mu$Jy) &($''\times''$) &(\deg) \\
    \hline
     PSR J1023 &10:23:47.689$\pm$0.001  &00:38:40.677$\pm$0.004 &103.7$\pm$3.6  &144.6$\pm$8.0  &(0.305$\pm$0.060)$\times$(0.205$\pm$0.031) &86$\pm$29  \\
     ALMA\,J102349.1  &10:23:49.131$\pm$0.002 &00:38:23.72$\pm$0.02 &95$\pm$15 &109$\pm$29 &<0.39$\times$0.22$^{\rm d}$ &\ldots\\
      \hline
    \end{tabular}
    \parbox{1.0\textwidth}{
    {\bf Notes.}
    $^{\rm a}$ Peak intensity.
    $^{\rm b}$ Integrated flux.
    $^{\rm c}$ FWHM obtained from 2D Gaussian fits.
    $^{\rm d}$ Component is a point source.
    }
     \label{tab:alma_res}
\end{footnotesize}
\end{table*}
%%%%%%%%%%%%%%%%%%%%%%%%%%%%%%%%%%%%%%%%%%%%%%

%%%%%%%%%%%%%%%%%%%%%%%%%%%%%%%%%%%%%%%%%%%%%%%%%%%%%%%%%%%%%%%%%%%%%%%%
%%%%%%%%%%%%%%%%%%%%%%%%%%%%%%%%%%%%%%%%%%%%%%%%%%%%%%%%%%%%%%%%%%%%%%%%

The 3.1 mm continuum emission from \src\ has a compact shape with some elongation towards the south-east (shown in the top right panel of Fig.\,\ref{fig:alma} using black contours). The millimetre continuum emission peaks $\sim$$0.32''$ south of the optical position. The deconvolved size obtained from the 2D Gaussian fit in the \texttt{robust=0} image is $0.305''\times0.205''$, which corresponds to approximately 418AU $\times$ 281AU ($\sim$$6.3\times10^{15}$~cm $\times$ $4.2\times10^{15}$~cm) at a distance of 1.37\,kpc (this is approximately 7$\times$10$^4$ times the orbital separation of the system; \citealt{archibald13}).
The emission from \srcmm\ also appears compact (see the bottom-right panel of Fig.\,\ref{fig:alma}).
The 2D Gaussian fit yields a point-like source with a full with at half maximum as large as $0.39''\times0.22''$ (see Table\,\ref{tab:alma_res}). The results of the searches for counterparts of this source at other wavelengths are reported in Appendix\,\ref{sec:appendix}.
%A discussion on the possible nature of this source is beyond the scope of this paper.

To determine the spectral index of \src\ within the ALMA bandwidth, we imaged the continuum emission from the lower and upper sidebands, which have central frequencies of 91.47886\,GHz and 103.5115\,GHz, respectively. The resulting spectral index is $-0.1\pm0.8$, which is consistent with previous reports for the average cm emission \citep{deller15,bogdanov18}.

To probe variability in the millimetre emission, we obtained images scan by scan (typically with a duration of $\sim$4--5 minutes), using a natural weighting.
We then computed the peak intensities and flux densities considering the emission above $3\sigma$ level.
The bottom panel of Fig.\,\ref{fig:timeseries2} shows the evolution of the flux density over time. This quantity was found to vary by a factor of $\sim$5 within the range $\sim$30--160\,$\mu$Jy on a timescale of about 5 minutes. In some cases, the flux density remained below the detection sensitivity, resulting in upper limits of the order of a few tens of microjanskys.
Figure\,\ref{fig:timeseries2} also shows the ALMA time series extracted using time intervals of variable length corresponding to the durations of the X-ray-mode episodes. Two flaring episodes can be clearly seen.
%Figure\,\ref{fig:high-low} shows the 3.1~mm continuum emission corresponding to the low X-ray mode (black contours) and high X-ray mode (white contours) overlaid with the natural weighting ALMA image obtained using the whole data set. As can be seen in Fig.\,\ref{fig:high-low} and Table\,\ref{tab:mode_properties}, the emission from the low X-ray mode is slightly more extended than the emission associated to the high X-ray mode. Besides, the peak emission corresponding to the high X-ray mode coincides with the peak position of the total 3.1~mm emission, whereas the peak from the low X-ray mode is slightly shifted to the south ($\sim0.062''$), but within the position accuracy of the mm emission.
We measured average flux densities of 89$\pm$13\,$\mu$Jy in the high mode and 134$\pm$23\,$\mu$Jy in the low mode, excluding the two flaring episodes (see Table\,\ref{tab:mode_properties}). 
Overall, the anti-correlated variability pattern appears to be more pronounced between the cm and X-ray emission than between the millimetre and X-ray emissions.

The integration times adopted above are effective for studying the variability of the millimetre emission on minute timescales, but may average out large enhancements of the millimetre flux on much shorter timescales (of the order of seconds). 
Therefore, we performed a dedicated search for short flaring episodes by extracting images of different durations using the entire dataset. This search resulted in the detection of the two above-mentioned episodes of bright millimetre emission (Fig.\,\ref{fig:timeseries_alma_flares}). No other remarkable events were detected. The first episode (hereafter `Flare 1') occurred on 2021 June 26 within the time range 22:19:29 -- 22:19:44 (UTC) at the switch from high mode to low mode and reached a peak flux density of 1.01$\pm$0.08\,mJy. The second event (`Flare 2') took place during the time range 23:48:29 -- 23:48:44 (UTC) and reached a peak flux density of 0.7$\pm$0.1\,mJy. 
The lack of continuous coverage in the millimetre band before and during the switch from the high to the low mode does not allow us to determine the exact epoch of the flare onset.
We note that these values refer to the flux at the peak of the flares and are therefore higher than those obtained in the X-ray mode-resolved analysis (see Fig.\,\ref{fig:timeseries2} and Table\,\ref{tab:flares}), where the emission is averaged over longer time intervals. We observed a significant increase in the peak intensity above the average level: a factor of $\sim$4 for Flare~1 and a factor of $\sim$7 for Flare~2. 

\begin{figure*}
    \centering
    \includegraphics[width=0.72\textwidth]{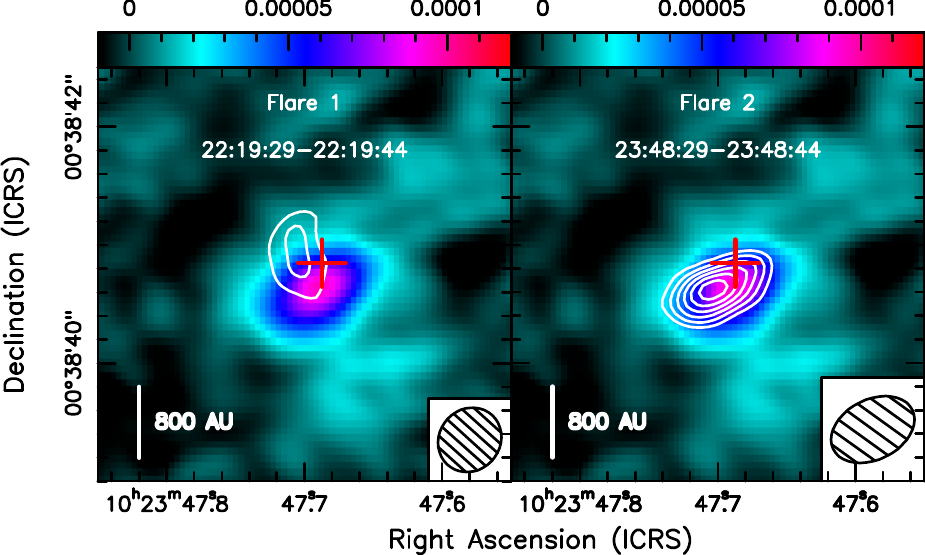} 
    \caption{Contour levels for the millimetre emission detected during 15 s time intervals at the peak of Flare~1 (left panel) and Flare~2 (right panel) shown with solid white lines overlaid on the naturally weighted ALMA 3.1~mm image extracted using the whole integration time (colour scale). These contour levels start at $3\sigma$ and increase in steps of $1\sigma$, where $\sigma$ is the rms of the image, $\sim$90\,$\mu$Jy\,beam$^{-1}$. Both panels display the same colour scale range in units of Jy\,beam$^{-1}$. Time intervals (in UTC) are displayed at the top of each panel. The red cross depicts the radio position of \src\ \citep{deller12}. The synthesised beams, with extension $0.696''\times0.540''$ and position angle P.A.$=-79.92^{\circ}$ for Flare~1 and $0.931''\times0.525''$ and P.A.$=-74.21^{\circ}$ for Flare~2, are shown at the bottom-right corner of each panel.}
    \label{fig:flares}
\end{figure*}

We observed a clear change in the morphology of the millimetre continuum emission during the two flaring episodes (see the white contours in Fig.\,\ref{fig:flares}). The 2D Gaussian fitting (see Table\,\ref{tab:flares}) indicates that the emission is elongated along the north-east to south-west direction during Flare~1 (roughly perpendicular to the ALMA 3.1~mm emission obtained with all the visibilities), while it is marginally resolved in only one direction during Flare~2. 
%This change in morphology suggests that the ejection pattern during flares is essentially stochastic.
The nominal rms absolute positional accuracy for standard ALMA observations can be estimated as $beam_{\rm FWHM} \times (0.9 \times {\rm S/N})^{-1}$ \footnote {See \url{https://almascience.eso.org/documents-and-tools/cycle10/alma-technical-handbook}, Sect.\,10.5.2}, where $beam_{\rm FWHM}$ is the full width at half maximum synthesised beam in arcsec and S/N is the signal-to-noise ratio of the image target. The positional accuracy for Flare~1 is $\simeq0.12-0.15''$, while for Flare~2 it is slightly smaller at $\simeq0.1''$. However, the actual absolute positional accuracy may be worse than the nominal values, potentially by a factor of 2 or more, depending on the atmospheric phase conditions during the observation. The offset between the ALMA peak positions and the radio position \citep{deller12} is $\simeq$0.4$''$ for Flare~1 and $\simeq$0.3$''$ for Flare~2. Given the stable weather conditions during the observation, it is possible that this offset is genuine for both flares. However, caution must be exercised, as this offset is comparable to the positional accuracy of ALMA.

%%%%%%%%%%%%%%%%%%%%%%%%%%%%%%%%%%%%%%%%%%%%%%%%%%%%%%%%%%%%%%%%%%%%%%%%
\begin{table*}
\footnotesize
\centering
\caption{
\label{tab:mode_properties}
X-ray and millimetre properties of \src\ during the periods of high and low X-ray modes.}
\begin{tabular}{ccccccc}
\hline\hline
X-ray mode      & $\Gamma$$^{\rm a}$    & $F_{\rm X}$$^{\rm b}$ & $I_{\nu}$     & $S_{\nu}$   & Deconvolved Size & P.A  \\  
                        &                                               & (10$^{-12}$~\flux)      & ($\mu$\Jy)    & ($\mu$Jy)   & $(''\times'')$   & (\grau) \\
\hline
High            & $1.66\pm0.03$                 & $13.8\pm0.2$          & $74\pm6$      & $89\pm13$    & (0.338$\pm$0.202)$\times$(0.250$\pm$0.044) &105$\pm$8\\
Low             & $1.7\pm0.1$                   & $2.1\pm0.2$               & $117\pm11$    & $134\pm23$  & (0.384$\pm$0.245)$\times$(0.260$\pm$0.125) &141$\pm$45\\
\hline   
\end{tabular}
\parbox{1.0\textwidth}{
{\bf Notes.}
$^{\rm a}$ Best-fitting photon index for the absorbed power-law model.
$^{\rm b}$ Unabsorbed flux over the 0.3--10\,keV energy band.
}
\end{table*}
%%%%%%%%%%%%%%%%%%%%%%%%%%%%%%%%%%%%%%%%%%%%%%%%%%%%%%%%%%%%%%%%%%%%%%%%

%%%%%%%%%%%%%%%%%%%%%%%%%%%%%%%%%%%%%%%%%%%%%%%%%%%%%%%%%%%%%%%%%%%%%%%%
%%%%%%%%%%%%%%%%%%%%%%%%%%%%%%%%%%%%%%%%

%%%%%%%%%%%%%%%%%%%%%%%%%%%%%%%%%%%%%%%%%%%%%%%%%%%%%%%%%%%%%%%%%%%%%%%%
\begin{table*}
\footnotesize
\caption{
\label{tab:flares}
Millimetre properties of \src\ during the two flaring episodes.}
\centering
\begin{tabular}{lcccccc}
\hline\hline
Flare & R.A (ICRS) & Dec. (ICRS) & $I_{\nu}$$^{\rm a}$ & $S_{\nu}$$^{\rm a}$ & Deconvolved Size & P.A \\
     &(hh:mm:ss) &(\deg:$'$:'') &($\mu$\Jy) &($\mu$Jy) &($''\times''$) &(\deg) \\
     \hline
Flare~1 &10:23:47.705$\pm$0.006   &00:38:40.92$\pm$0.11   &426$\pm$26   &1006$\pm$84   & (0.96$\pm$0.89)$\times$(0.40$\pm$0.10)   &13.3$\pm$6.9   \\
Flare~2 &10:23:47.702$\pm$0.002   &00:38:40.63$\pm$0.01   &701$\pm$62   &675$\pm$114   &\ldots$^{\mathrm{b}}$   &\ldots$^{\mathrm{b}}$   \\
\hline   
\end{tabular}
\parbox{1.0\textwidth}{{\bf Notes.}
$^{\rm a}$ These values are measured at the peak of the flares over time intervals of length 15\,s and are therefore higher than those obtained in the X-ray mode-resolved analysis and plotted in Fig.\,\ref{fig:timeseries2}, where the emission is averaged over longer time intervals. 
$^{\rm b}$ Source marginally resolved in only one direction.}
\end{table*}
%%%%%%%%%%%%%%%%%%%%%%%%%%%%%%%%%%%%%%%%%%%%%%%%%%%%%%%%%%%%%%%%%%%%%%%%

\subsection{Radio emission}
\subsubsection{Continuum emission}
\label{sec:radio_results}
The radio light curve extracted from VLA data is shown in Fig.\,\ref{fig:timeseries1}. J1023  is not detected most of the time, especially during the high mode intervals. However, it is detected during almost all episodes of low mode (here the uncertainties on the flux densities are reported at the 1$\sigma$ confidence level). The average radio flux densities measured separately in the high and low mode are reported in Table\,\ref{tab:SED}. These values are consistent with those reported by \cite{bogdanov18} within the uncertainties. This indicates that the radio flux densities measured separately in the two modes are remarkably stable over timescales of years.

\subsubsection{Non-detection of pulsed emission}
\label{sec:radioUL}
The limit on the radio pulsed flux density derived from the FAST observations is a factor of $>$5000 smaller than the flux density measured at similar frequencies when \src\ was active as a radio pulsar \citep{Archibald2009} and a factor of a few tens more constraining than the limits previously evaluated (again at similar frequencies) as soon as \src\ transitioned to the sub-luminous X-ray state \citep{stappers14,patruno14}. The limit on the radio pulsed flux density in the high mode is also at least four orders of magnitude smaller than the value predicted by extrapolating the best-fitting power-law model for the SEDs of the optical-to-X-ray pulsed emission \citep{papitto19,miravalzanon22}. 

\subsection{The low-mode ingress and egress timescales}
\label{sec:lowmode_times}
We first examined the low mode ingress and egress times in the \xmm\ observation. For each low-mode episode, we used a simple model for the count rate light curve consisting of four components: a constant level for the high mode (allowing for a different count rate at the entrance and at the exit of the low-mode episode), a linear ingress in the low mode, a constant count rate in the low mode, and a linear egress. This model is similar in some aspects to what is used to model eclipses in binary stars or planetary transits. While we do not expect to accurately describe each low-mode episode individually, we consider the results in a statistical sense.

We applied this model to each low-mode episode in the \xmm/EPIC count rate light curve, which was binned at 10\,s, excluding the initial flaring mode of the observation. We performed the same procedure on the NICER data from the second night, using again a 10 s time bin. We fit the model to 56 low-mode episodes for \xmm\ and 10 for NICER. 

\begin{figure*}
\begin{center}
\includegraphics[width=1.0\textwidth]{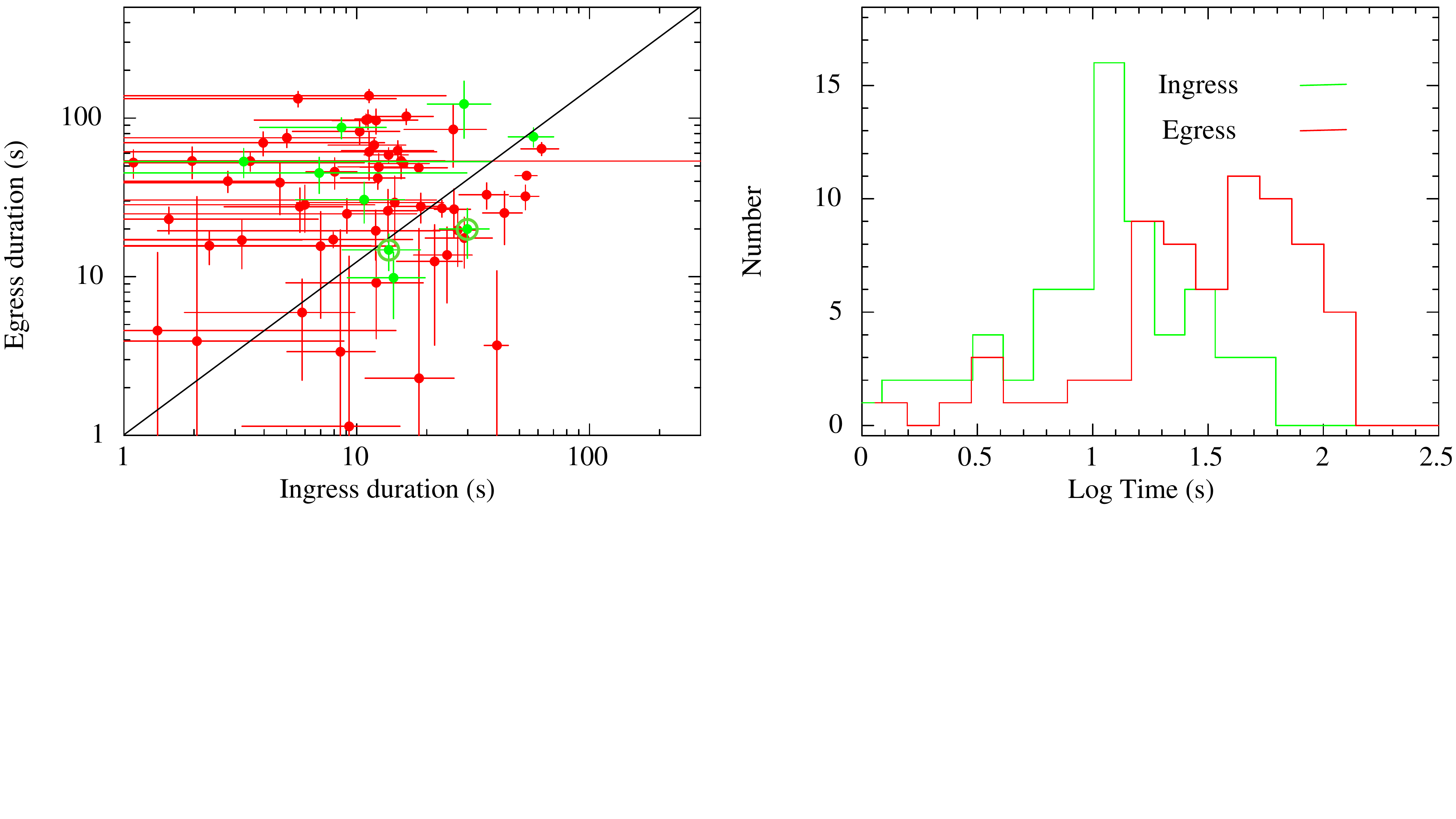}
\vspace{-4.2cm}
\caption{Results of the statistical analysis of the X-ray mode switching timescales. \textit{Left}: Duration of the low-to-high mode switch (egress) as a function of the duration of the high-to-low mode switch (ingress) computed using data from \xmm/EPIC (in red) and NICER (green). The circled green points represent the values for the ingress associated with the two millimetre flares detected with ALMA and the subsequent egress. The solid black line indicates where the two quantities are equal. \textit{Right}: Distribution of the low-mode ingress (in green) and egress (in red) times in logarithmic scale from \xmm\ and NICER data.} 
\label{fig:ingress_egress}
\end{center}
\end{figure*}

As shown in Fig.\,\ref{fig:ingress_egress}, the ingress time is shorter than the egress time from the low mode ($\sim 80\%$). Interestingly, we observe that the two low-mode instances detected in the NICER data and coinciding with short-duration flares detected by ALMA both have comparable ingress and egress times (see the left panel of Fig.\,\ref{fig:ingress_egress}). This is uncommon and may be related to the detectability of flares at millimetre wavelengths. The distributions of \xmm\ low-mode ingress and egress times are shown in the right panel of Fig.\,\ref{fig:ingress_egress}. The average logarithmic ingress time is $(10.4\pm6.9)$\,s, which is consistent with the time binning. The average logarithmic egress time is $(30.6\pm8.2)$\,s.  
We also investigated whether there was any relationship between ingress time, egress time, duration of the low mode, count rate at the beginning and end of the low mode, and count rate in the low mode. However, we did not find any statistically significant correlations.

Although the model we used to describe the low-mode episodes is simple, it is clear that the duration of the ingress time in the low mode is short and consistent with the adopted time binning, and the egress time from the low mode is measurable and longer.

%%%%%%%%%%%%%%%%%%%%%%%%%%%%%%%%%%%%%%%%%%%%%%%%%%%%%
%%%%%%%%%%%%%%%%%%%%%%%%%%%%%%%%%%%%%%%%%%%%%%%%%%%%%

\section{Discussion}
\label{sec:discussion}

\subsection{The physical picture}
\label{sec:scenario}
\begin{figure}
\begin{center}
\includegraphics[width=0.49\textwidth]{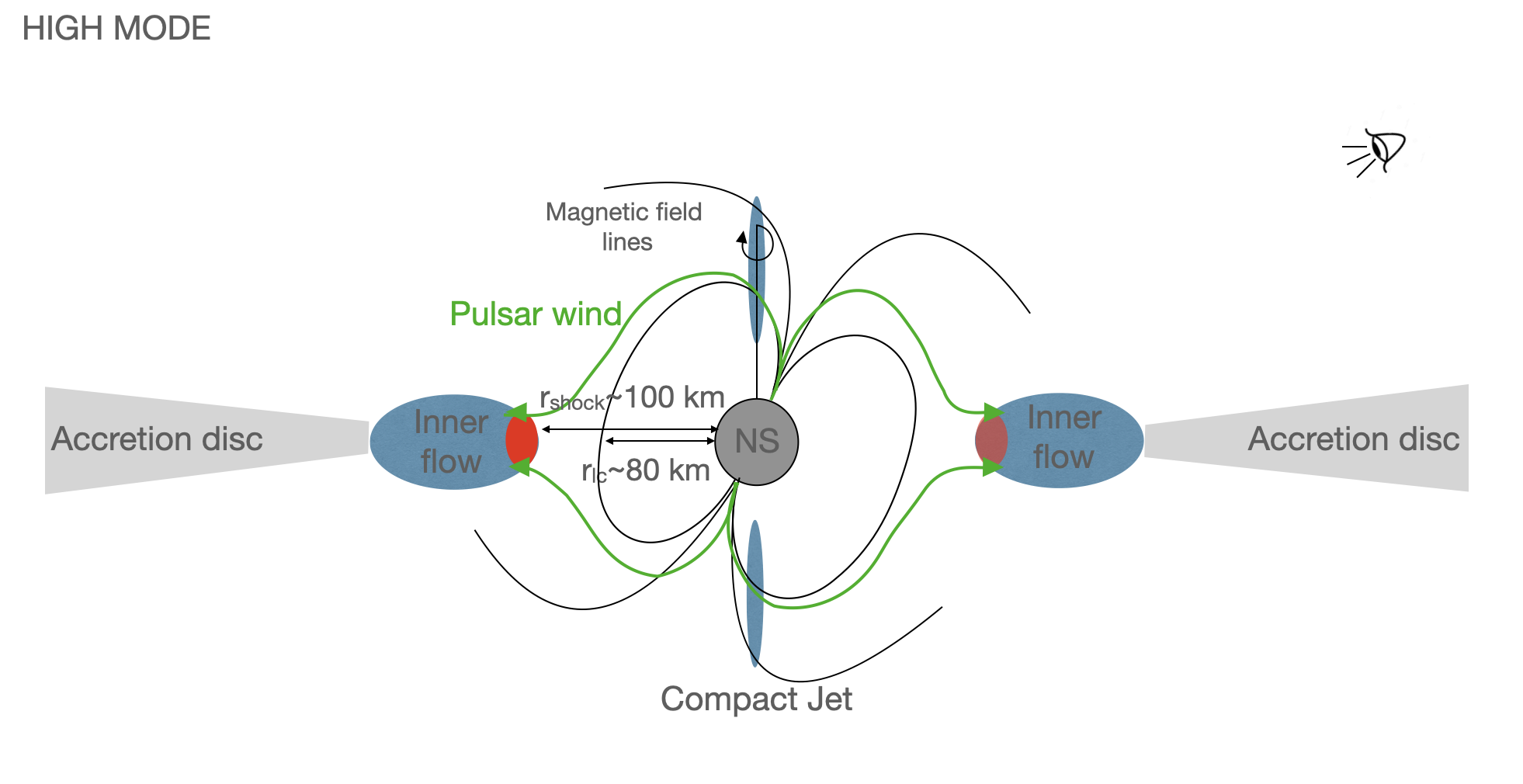}
\includegraphics[width=0.49\textwidth]{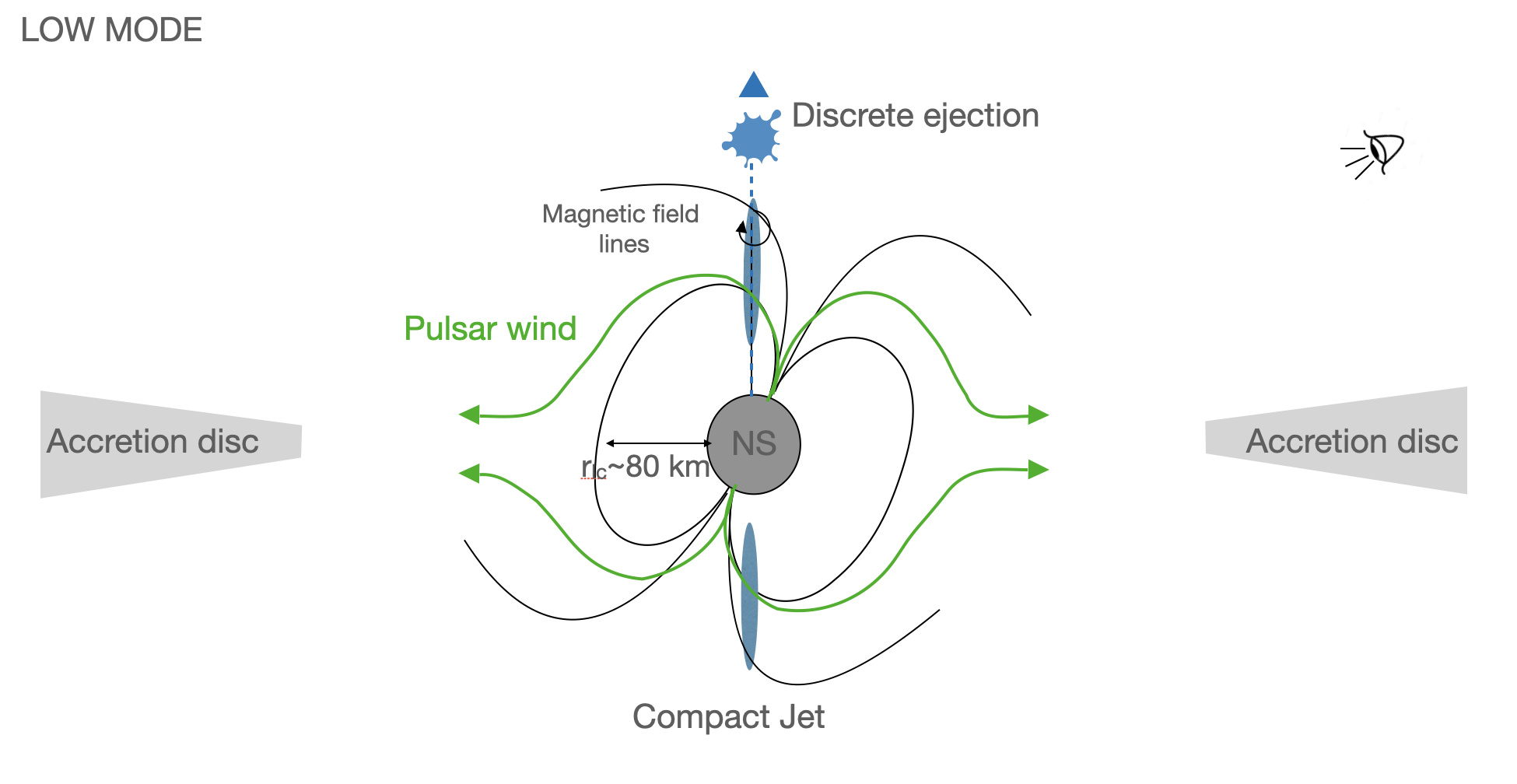}
\caption{
Schematic visual representation of the evolution of the inner flow into an outflow at the switch from the high to the low X-ray emission mode. When the system is in the high mode (top panel), a small-size inner flow is present, together with a faint steady jet that is launched along the pulsar rotational axis and gives rise to the observed low-level radio and millimetre emission. As the pulsar rotates, the pulsar wind (marked with solid green lines) wobbles around the equatorial plane (see e.g. \citealt{bogovalov99}) and shocks off the electrons in the inner flow at two opposite sides (red spots) at a distance that is slightly larger than the light cylinder radius ($\simeq$80\,km).
At each pulsar rotation, synchrotron emission at the shock at X-ray, UV, and optical frequencies is modulated at the spin period at one side (bright red spot), while it is absorbed by material in the inner flow at the other side \citep[light red spot;][]{papitto19}. 
When the system makes a switch to the low mode (bottom panel), a bright discrete ejection is launched along the pulsar rotational axis on top of the compact jet, the inner flow disappears, and the shock emission is quenched. For displaying purposes, only the inner regions of the compact jet are represented in both panels. 
} 
\label{fig:sketch}
\end{center}
\end{figure}

Our campaign, along with previous findings (see e.g. \citealt{ambrosino17,bogdanov18,papitto19,miravalzanon22,illiano23} and references therein), allows us to draw a comprehensive physical picture of the high-low mode switches in \src, which involves a permanently active rotation-powered pulsar, an accretion disc and episodes of discrete mass ejection on top of a compact jet. 
In the following, we outline the physical scenario while in Sect.\,\ref{sec:sed} we describe in detail the SED modelling in support of our scenario.
Figure\,\ref{fig:sketch} shows a visual representation of our scenario. We do not discuss the flaring activity here as it was not covered over a wide energy range.

\begin{figure*}
    \centering
    \includegraphics[width=1.\textwidth]{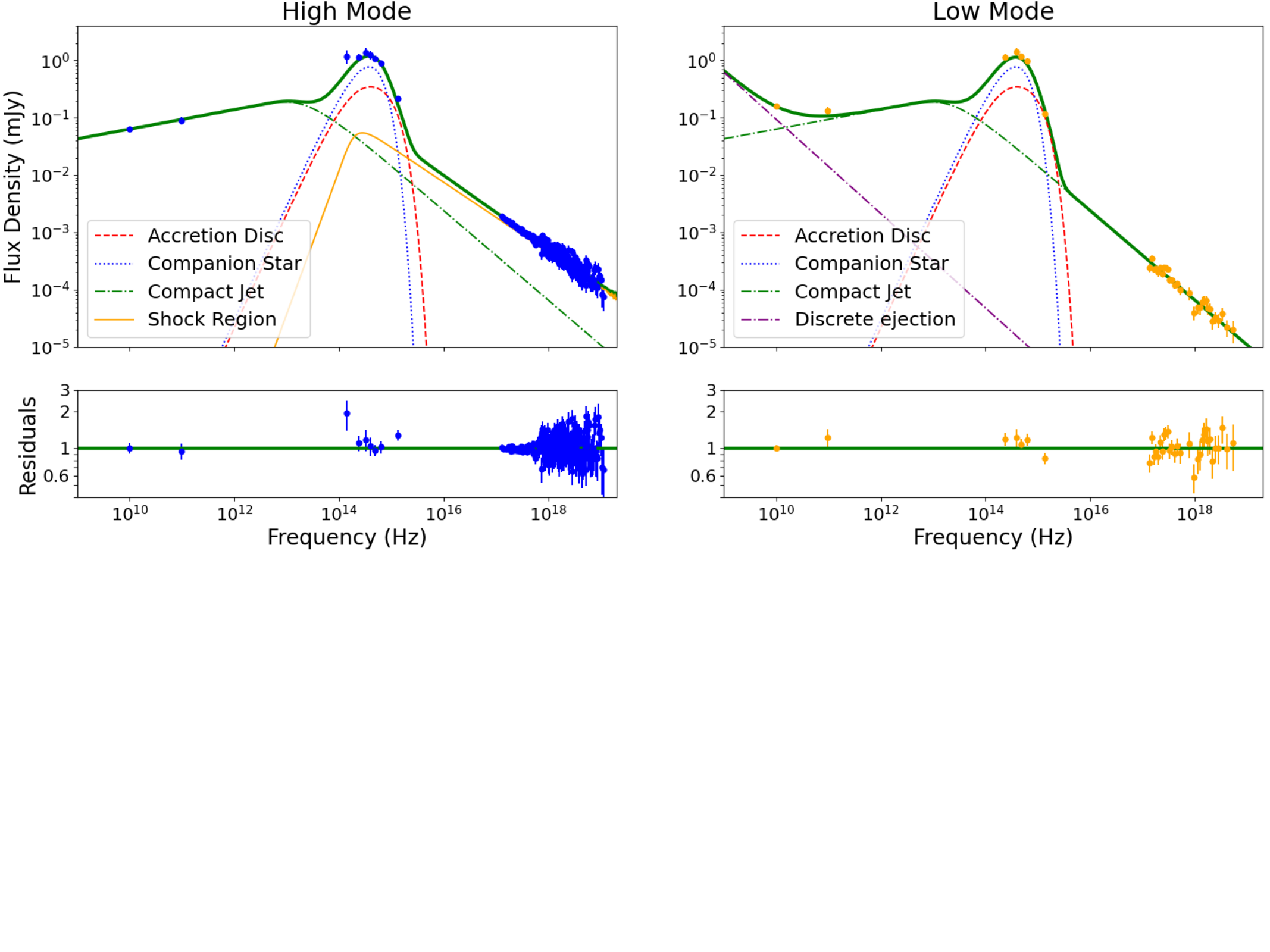}
    \vspace{-6cm}
    \caption{Broadband (radio to X-rays) SEDs of \src\ during the high and low modes, along with the best-fitting models. Both SEDs were corrected for extinction effects. The red-dashed, blue-dotted, purple-dashed-dotted and orange solid lines represent the contributions of the accretion disc, companion star, discrete ejection and shock region, respectively. The green dashed line represents the compact jet, whose contribution does not vary between the two modes. The green solid line is the sum of all contributions in both SEDs. The ratio between the data points and the best-fitting model is shown at the bottom of each panel.}
    \label{fig:SED}
\end{figure*}

We assume that during the high mode the accretion flow is kept just beyond the light cylinder radius ($\simeq$80\,km) by the radiation pressure of the particle wind from an active rotation-powered pulsar \citep{papitto19}.
This causes the innermost region of the truncated, geometrically thin accretion disc to be replaced by a radiatively inefficient, geometrically thick flow, as expected for low-accretion rates \citep{narayan94}. In this configuration, synchrotron radiation at a shock front that forms where the pulsar wind interacts with the inner flow produces the bulk of the X-ray emission as well as X-ray, UV, and optical pulsed emission \citep{papitto19, veledina19}. 
The collision of high-energy charged particles from the pulsar wind with the in-flowing matter in the disc increases the ionisation level of the disc. As a result, the in-flowing matter is drawn into the pulsar magnetic field and accelerated, producing a compact jet of plasma that streams out. According to our SED modelling, the corresponding jet synchrotron spectrum is optically thick up to the mid-infrared band and breaks to optically thin above a frequency of $(2.4^{+1.9}_{-1.0})\times 10^{13}$\,Hz, giving only a small contribution at X-ray energies ($\sim$8--18\% in the 0.3--10\,keV band; see Fig.\,\ref{fig:SED} and Table\,\ref{Tab:fit_res}).

The switch from the high to the low mode can be connected to the occurrence of short millimetre flares. There is strong evidence that at least one such flare is directly involved in triggering the switch (see Fig.\,\ref{fig:timeseries2}). We interpret these flares as the fingerprint of a discrete mass ejection that marks the removal of the inner flow (see Sect.\,\ref{sec:mmflares}), similar to what has been observed in other black-hole and NS XRBs (e.g. Cyg~X-3, Circinus~X-1, V404~Cygni, MAXI~J1535-571, and MAXI~J1820+091; \citealt{egron21,fender98,fender23, RussellT2019, Homan2020}), albeit in different accretion regimes. As soon as the inner flow is expelled in the discrete ejection, synchrotron emission from the shock no longer takes place, resulting in a drop of the X-ray flux and the quenching of X-ray, UV and optical pulsations. The system enters the low mode.

During the low mode, the pulsar wind can still penetrate the accretion disc, collide with the in-flowing matter, and launch the compact jet, contributing to the multi-band emission in the same way as in the high mode. The X-ray emission is likely to be (almost) entirely from optically thin synchrotron radiation from this jet (Fig.\,\ref{fig:SED}, right panel), with a power-law index of $\alpha = -0.78\pm0.04$ (where the flux density $F_{\nu}$ scales with the observing frequency $\nu$ as $F_{\nu}\propto \nu^{\alpha}$; see Table\,\ref{Tab:fit_res}). At the same time, the emission associated with the discrete ejection progressively shifts towards lower frequencies as the ejected material expands adiabatically during the low mode and quickly evolves from optically thick to optically thin \citep{bogdanov18}. This emission component has a steep spectrum ($\alpha = -0.8_{-0.4}^{+0.3}$) and provides an additional flux contribution above that of the underlying optically thick emission from the compact jet at frequencies below the millimetre band (see Figs.\,\ref{fig:timeseries1} and \ref{fig:timeseries2}).

Eventually, the flow from the disc starts to refill the inner regions just outside the NS light cylinder on a thermal timescale (tens of seconds) due to its advection-dominated nature. The refilling of the disc re-establishes the region of shock with the pulsar wind, leading to increased emission and pulsations at  X-ray, UV, and optical frequencies \citep{papitto19,miravalzanon22,illiano23}. The system thus makes a switch back to the high mode.

As shown in Sect.\,\ref{sec:lowmode_times}, the low mode ingress and egress times show a statistically significant difference, with the average egress time being longer than the average ingress time. 
In our picture, the ingress in the low mode is caused by the ejection of the innermost regions of the disc. This will occur on a dynamic timescale.
We can estimate the distance at which the innermost edge of the disc in \src\ must recede away from the NS at the high-to-low mode switch by assuming that most of the X-ray emission in the low mode is due to the jet and hence imposing that the luminosity of the synchrotron emission at the shock front is at most, say, 10\% of that observed in the low mode. This gives a radial distance of $r_{\rm low} = [L_{\rm sd}/(0.1L_{\rm X,low})]^{0.5}r_{\rm LC} \approx20\,r_{\rm LC} \approx$1600\,km ($L_{\rm sd}=4.43\times10^{34}$\,\lum\ is the spin-down luminosity; \citealt{archibald13}). This value should be considered as a lower limit: at larger radii, the X-ray luminosity due to the shock would be even smaller. The dynamic timescale is $\sqrt{r^3/(GM)}\sim 0.1$ s at $20\,r_{\rm LC}$. Observations from \xmm\ and NICER show that current X-ray instruments are capable of probing mode switching in \src\ on timescales as short as $\sim$10\,s. In the future, proposed X-ray missions that offer instruments with much larger collecting areas, high throughput, and excellent time resolution, such as the New Advanced Telescope for High-ENergy Astrophysics (\emph{NewAthena}; \citealt{nandra13}), the enhanced X-ray Timing and Polarimetry mission (eXTP; \citealt{Zhang2016}) and the Spectroscopic Time-Resolving Observatory for Broadband Energy X-rays (STROBE-X; \citealt{ray19}), would be able to sample the switch on a timescale of $<$10\,s and test our predictions.

The ejection and replenishment of the inner flow in \src\ during the switches from high to low mode and from low to high mode are reminiscent of the ejection of the X-ray corona into a jet and subsequent replenishment observed in the Galactic microquasar GRS\,1915$+$105 \citep{mendez22}.
A previous study has demonstrated that, in the case of GRS\,1915$+$105, the outer disc steadily refills the central part of the system on a viscous timescale \citep{Belloni1997}. The rise time is the duration it takes for the heating front to propagate through the central disc. By assuming a subsonic velocity for the heating front, with the sound speed $c_S$ calculated in the cool ($\lesssim 6\times 10^6$\, K) disc state, we can estimate the rise time to be $r_{\rm low}/c_S\approx 20$\,s for \src.

In the above picture, we have assumed that the emission from the compact jet remains relatively constant over time, but it is also possible that the jet could be partially disrupted by the discrete ejection of plasma at the switch from high to low mode. However, if the ejection is not energetic enough to overcome the magnetic fields and pressures within the jet, or if it is not dense or well aligned enough to effectively interact with the jet, then the jet would quickly re-establish itself and appear as if it had never turned off. In either case, the jet would still contribute the same amount of flux in both modes, especially in the radio band where the emission is produced at greater distances from the NS (of the order of light-minutes to light-hours; \citealt{Chaty2011}) and therefore any changes are on much longer timescales than the mode switching.

%%%%%%%%%%%%%%%%%%%%%%%%%%%%%%%%%%%%%%%%%%%%%%%%%%%%%
%%%%%%%%%%%%%%%%%%%%%%%%%%%%%%%%%%%%%%%%%%%%%%%%%%%%%

\subsubsection{Millimetre flares as the fingerprint of discrete ejecta}
\label{sec:mmflares}
Starting from the timescales of the first millimetre flare that occurred on the second night (which has more accurate start and end times than the second flare) and the typical timescales of the radio flares that occurred on the first night, we attempted to reproduce the light curves of similar flares by assuming they are caused by the motion and expansion of a blob of plasma accelerated near the pulsar. Following the work of \cite{fender19}, we assumed that a typical discrete ejection expansion is well described by the van der Laan model \citep{vanderlaan1966}\footnote{The calculations we made are based on \cite{vanderlaan1966} and are described in a Jupyter-notebook available at \url{https://github.com/robfender/ThunderBooks/blob/master/Basics.ipynb}}. Since no flare in this study was observed simultaneously at two different frequencies, we are unable to properly constrain the model parameters. However, we are able to check whether we could reproduce flares with peak flux densities and durations that are consistent with what we observed. We assumed that the blob is launched a few hundred kilometres away from the NS with a mildly relativistic velocity ($v\approx0.05$\,$c$), and expands linearly over time. We also assumed a magnetic field that scales with the distance from the compact object as $B \propto R^{-1}$, with a magnetic field at the launching site of the order of 10$^5$\,G (see Eq.\,6 in \citealt{papitto19} for an expression of the post-shock magnetic field of an isotropic pulsar wind). Additionally, we considered an electron number density of 10$^{11}$--10$^{12}$\,cm$^{-3}$, which is consistent with the lower limit estimated from high-resolution X-ray spectroscopy \citep{cotizelati18a}. Our calculations show that we can reproduce flares that last for a few seconds and reach a peak flux density of the order of 1\,mJy at the ALMA Band-3 centroid frequency. Furthermore, these flares would result in flares that last for a few tens of seconds and reach a peak flux density of the order of 0.2\,mJy at the VLA Band-X centroid frequency, as observed. These results are consistent with the predictions of the van der Laan model for an expanding blob of plasma.

Although we cannot firmly prove the true origin of the flares, we can ascribe the non-detection of flares in five out of the seven high-to-low-mode switches covered by ALMA to the variation in the intrinsic properties of the ejected material. These include the size, density, temperature, electron energy distribution, and opening angle of the ejected material. In general, smaller, less dense, and cooler ejecta that travel at slower bulk speeds tend to produce fainter millimetre emission (see e.g. \citealt{tetarenko17,wood21}). Additionally, the distribution of the ejected material along the line of sight can also affect the detectability of millimetre flares through absorption and scattering effects (this is likely the main reason for the non-detection of pulsed radio emission as well; see Sect.\,\ref{sec:mass_transfer_rate}). Overall, a combination of these factors can explain the varying emission properties observed at millimetre and radio frequencies during (and following) different high-to-low-mode switches. These variations are likely to affect the time it takes for the inner flow to restore itself and produce the shock emission, resulting in the observed differences in duration among the individual low-mode episodes.

\subsubsection{The non-detection of radio pulsations} 
\label{sec:mass_transfer_rate}
The non-detection of radio pulsations could be due to `intrinsic' phenomena within the pulsar magnetosphere (i.e. irrespective of the presence of external drivers such as accreting material) such as intermittency and/or nulling (see e.g. \citealt{weng22} for recent similar considerations for the system LS I $+$61 303). However, these phenomena were not observed during extensive radio timing campaigns when the system was in the radio pulsar state (e.g. \citealt{archibald13,stappers14}), so they are unlikely to be the cause of the non-detection. Alternatively, the non-detection of pulsations may be due to the presence of matter enshrouding the system (see also \citealt{cotizelati14,stappers14}). In this scenario, we can estimate a lower limit on the mass transfer rate from the companion star by requiring that the optical depth due to free-free absorption is larger than unity throughout the duration of the FAST observation. The analytical expression for the free-free optical depth is given by

%%%%%%%%%%%%%%%%%%%%%%%%%%%%%%%%%%%%%%%%%%%%%%%%%%%%%%%%%%%%%%%%%%%%%%%%%%%%%%%%%%%%%%%%%%%%%%%%%%%%%%%%%%%%%%%%%%%%%%%%%%
\begin{equation}
\begin{split}
\MoveEqLeft
\tau_{\rm ff,1.25\,GHz}\simeq2.126\times10^{22}\times\\ &\times \frac{\dot{m}_{\rm C}^2 (M_{\rm NS}+M_{\rm C})^2 (X+0.5Y)^2 (9.157+2\ln T_4)}{T_4^{3/2}v_8^2 P_{\rm orb}^2 (M_{\rm NS}+0.538M_{\rm C})^3}~,
\label{eq:tauff}
\end{split}
\end{equation}where $\dot{m}_{\rm C}$ is the mass transfer rate in units of $M_{\odot}$\,yr$^{-1}$, $M_{\rm NS}$ and $M_{\rm C}$ are the mass of the NS and the companion star, respectively, $X$ and $Y$ are the mass fraction of hydrogen and helium, respectively, $T_4$ is the plasma temperature in units of $10^4$\,K, $v_8$ is the velocity of the particle wind in units of $10^8$\,m\,s$^{-1}$ and $P_{\rm orb}$ is the orbital period in hours \citep{burgay03a,iacolina09}. We assume $M_{\rm NS}$ = 1.65\,$M_{\odot}$, $M_{\rm C}$= 0.24\,$M_{\odot}$, $P_{\rm orb}$=4.75\,h \citep{Archibald2009}, $v_8$=1, $X$=0.7 and $Y$=0.3\footnote{We note that our assumption on $X$ and $Y$ should be regarded as a zeroth-order approximation since a recent study by \cite{shahbaz22} has shown that the chemical abundances of the companion star of \src\ are different from those found in most XRBs. However, since $\dot{m}_{\rm C} \propto (X+0.5Y)^{-1}$, the lower limit on $\dot{m}_{\rm C}$ only increases by a factor of a few if smaller values of $X$ and $Y$ are assumed.}, and $v_8$=1. 
In order to have $\tau_{\rm ff,1.25\,GHz}\simeq 1$, and therefore not detect the radio signal due to free-free scattering, we estimate $\dot{m}_{\rm C}\gtrsim10^{-11}$\,$M_{\odot}$\,yr$^{-1}\approx6\times10^{14}$\,g\,s$^{-1}$. This value is a factor of $\gtrsim$10--60 larger than the mass accretion rate estimated based on a systematic analysis of the X-ray aperiodic variability of \src\ in the sub-luminous X-ray state \citep{linares22}. 
This implies that a large fraction of the accreting mass does not reach the compact object's surface and is instead ejected from the system. 
It is worth mentioning that the predicted mass transfer rate for \src, based on the transfer of mass through the loss of orbital angular momentum, is $\sim 10^{-11}M_{\odot}\rm \, yr^{-1}$ \citep{verbunt93}.

\subsection{Spectral energy distributions}
\label{sec:sed}

\subsubsection{Data preparation}
We extracted the SEDs separately in the high and low mode and fit them using a model that reflects the physical scenario proposed in Sect.\,\ref{sec:scenario}.

The NIR, optical and UV data obtained on Night\,2 were de-reddened assuming $A_V=0.17\pm 0.01$\,mag, calculated from the \xmm\ estimate of $N_{\rm H}$ using the $N_{\rm H}$/$A_V$ relation reported by \cite{Foight2016}. The absorption coefficients in the other bands were estimated using the equations reported by \cite{cardelli89}, and are listed in the last column of Table\,\ref{tab:SED}.

For the low-mode SED, we considered optical data acquired between MJD  (Modified Julian Day) 59392.01085 and 59392.01434, totaling three consecutive 60 s integration images in each band; the $J$-band flux is the average of the fluxes measured with NTT/SOFI while the source was in the low mode, according to our \xmm\ monitoring; the UV (uvm2) flux was measured on MJD 59392.01302; radio flux densities are taken from the VLA observation from 2021 June 3--4; the millimetre flux density was derived from stacking all the ALMA images acquired during the low modes. No $K$-band observations have been acquired during the low mode. The unabsorbed X-ray flux was calculated as the average flux during all periods of low mode.

For the high-mode SED, we considered $K$-band data acquired between MJD 59392.00237 and 59392.00289, and then from MJD 59392.00593 to 59392.00832 (i.e. all the images in which \src\ is detected, $15\times15$\,s); the $J$-band flux is the average of all the fluxes measured with NTT/SOFI while the source was in the high mode, according to our XMM-Newton monitoring; for the optical $griz$ bands, we considered one 60 s image for each band, at MJD 59392.00528; the UV (uvm2) flux was acquired on MJD 59392.00413; radio flux densities are taken from the VLA observation on 2021 June 3--4; the millimetre flux density was derived after stacking all the ALMA images acquired during the high mode episodes. The unabsorbed X-ray flux was calculated as the average flux during all periods of high mode.

%%%%%%%%%%%%%%%%%%%%%%%%%%%%%%%%%%%%%%%%%%%%%%%%%%%%%%%%%%%%%%%%%%%%%%%%
\begin{table}[]
    \centering
    \footnotesize
        \caption{Central frequencies ($\nu_c$), flux densities corrected for interstellar absorption, and the adopted absorption coefficients ($A_{\nu}$) used for the SEDs.}
    \begin{tabular}{ccccc}
    \hline
    \hline
         Band & $\nu_c$ (Hz) & \multicolumn2c{Flux Density (mJy)} & $A_{\nu}$ (mag) \\
                &              &    Low mode        &   High mode       &       \\
         \hline
         Radio & $1.00\times 10^{10}$ & $0.158\pm 0.009$    & $0.064\pm0.006$   & --  \\
           mm & $9.75\times 10^{10}$ & $0.133\pm0.023$ & $0.089\pm0.013$  & --   \\
          $K$ & $1.40\times 10^{14}$ &-- & $1.17\pm0.32$& $0.02\pm 0.03$ \\
         $J$ & $2.39\times 10^{14}$ & $1.14\pm0.13$ & $1.12\pm0.16$ & $0.05\pm 0.03$ \\
         $z$ & $3.28\times 10^{14}$ &$1.35\pm 0.28$ &$1.34\pm0.27$ & $0.08 \pm 0.01$\\
         $i$ & $3.93\times 10^{14}$ &$ 1.32\pm 0.23$ & $ 1.20\pm0.21$& $0.11\pm 0.01$ \\
          $r$ & $4.81\times 10^{14}$ &$ 1.09\pm0.11$& $ 1.03\pm0.11$& $0.14\pm 0.01$ \\
           $g$ & $6.29\times 10^{14}$ & $ 0.87\pm0.10$& $ 0.84\pm0.10$& $0.20\pm 0.01$ \\
          $uvm2$ & $1.34\times 10^{15}$ & $ 0.10 \pm 0.01$ & $0.18\pm0.02$& $0.52\pm 0.01$\\
          \hline
    \end{tabular}\\
    \label{tab:SED}
%{\bf Notes.}
%The X-ray data reported in the SEDs will be published electronically.
\end{table}

%%%%%%%%%%%%%%%%%%%%%%%%%%%%%%%%%%%%%%%%%%%%%%%%%%%%%%%%%%%%%%%%%%%%%%%%

\subsubsection{Model setup}
\textit{Low mode.} To perform the fit to the SED, we used a model comprising a blackbody of an irradiated star and a multi-colour accretion disc, together with a broken power law to describe the optically thick and thin components of the synchrotron emission from the compact jet. In addition, we considered a power law to model the optically thin emission from the discrete ejecta at all frequencies.
For the stellar component, we considered an irradiated star model (see Eqs.\,8 and 9 of \citealt{chakrabarty98}) where the surface temperature of the star ($T_{*}$) is allowed to vary and the distance to \src\ ($D$), the radius of the companion star ($R_{\rm c}$), the binary separation ($a$), the irradiation luminosity ($L_{\rm irr}$) and the albedo of the star ($\eta_{*}$) are fixed to known (or reasonable) values: $D=1368$\,pc \citep{deller12}, $R_{\rm c}=0.43\, R_{\odot}$ \citep{Archibald2009}, $\eta_{*}=0.1$, and $a=[G(M_{\rm NS}+M_{\rm C})P_{\rm orb}^2/(4\pi)^2]^{1/3}$, where $M_{\rm NS}=1.65M_{\odot}$ is the NS mass, $M_{\rm C}=0.24M_{\odot}$ is the mass of the companion star, $P_{\rm orb}=4.75$\,hrs is the binary orbital period and $G$ is the gravitational constant. 
$L_{\rm irr}$ was fixed to $6.5\times10^{33}$\,\lum, the value estimated by \cite{shahbaz19} from the difference of the temperatures of the illuminated and dark sides of the companion star of \src. This value corresponds to 14\% of the pulsar spin-down luminosity \citep{archibald13}, and is of the same order of magnitude as the X-ray and gamma-ray luminosities in the sub-luminous X-ray state.  
For the multi-colour accretion disc model (Eqs.\,10--15 by \citealt{chakrabarty98}), we allowed the inner radius of the accretion disc where the optical/UV emission becomes relevant ($r_{\rm in, opt/UV}$) to vary, while we fixed the X-ray albedo of the disc to 0.95 \citep{chakrabarty98} and the mass transfer rate to $10^{-11}M_{\odot}\rm \, yr^{-1}$. This is the value predicted for an XRB system with a short orbital period hosting a main-sequence star and transferring mass through loss of angular momentum \citep{verbunt93}, and is consistent with the lower limit estimated from the non-detection of radio pulsations (Sect.\,\ref{sec:FAST}).
The broken power-law component has the normalisation $F_0$, the two slopes $\alpha_1$ and $\alpha_2$ and the break frequency $\nu_{\rm break}$ as free parameters; the power law associated with the emission from the discrete ejecta has the normalisation $F_{\rm thin\, ,10}$ (calculated at a frequency of 10\,GHz) and the slope $\alpha_{\rm thin}$ as free parameters.

%%%%%%%%%%%%%%%%%%%%%%%%%%%%%%%%%%%%%%%%%%%%%%%%%%%%%%%%%%%%%%%%%%%%%%%%
%%%%%%%%%%%%%%%%%%%%%%%%%%%%%%%%%%%%%%%%%%%%%%%%%%%%%%%%%%%%%%%%%%%%%%%%

%%%%%%%%%%%%%%%%%%%%%%%%%%%%%%%%%%%%%%%%%%%%%%%%%%%%%%%%%%%%%%%%%%%%%%%%
\begin{table*}[]
    \centering
        \caption{Results of the model fit of the SED in low and high modes.}
        \begin{tabular}{cllrc}
    \hline \hline
Component             & Parameter & \multicolumn2c{Posterior} & Prior bounds  \\  \cline{3-4}
                 &                 &    Low mode   &   High mode  &       \\ \hline \vspace{0.3cm}
Irradiated star   & $T_{*}$ (K) &\multicolumn{2}{c}{$6141.05^{+32.91}_{-30.37}$}&$T_{*} \sim \mathcal{N}(6128, 33)^{\ast}$\\  \vspace{0.35cm}
Accretion disc    & Log [$r_{\rm in,opt/UV}$ / 1~cm] & \multicolumn{2}{c}{$9.49\pm 0.02$}& $\rm Log\,r_{\rm in,opt/UV}\sim \mathcal{U}(7, 12)$\\   \vspace{0.05cm}
Compact jet       & Log [$\nu_{\rm break}$ / 1~Hz] & \multicolumn{2}{c}{$13.40\pm 0.26$ } &$\rm Log\, \nu_{\rm break} \sim \mathcal{U} (10,14)$\\ \vspace{0.05cm}
                & $\alpha_1$ & \multicolumn{2}{c}{$0.18\pm 0.06$} &$\alpha_1\sim \mathcal{N}(0.2,0.2)^{\dagger}$\\ \vspace{0.05cm}
                & $\alpha_2$ & \multicolumn{2}{c}{$-0.78\pm 0.04$} &$\alpha_2\sim\mathcal{N}(-0.7, 0.2)^{\dagger}$\\ \vspace{0.35cm}
                & Log [$F_0$ / 1~mJy] & \multicolumn{2}{c}{$-0.59_{-0.17}^{+0.15}$} &$\rm Log \, F_{0} \sim \mathcal{U}(-2, 0)$\\   \vspace{0.05cm}
Shock emission  & Log [$\nu_{\rm sync}$ / 1~Hz] &-- &  $14.34^{+0.44}_{-0.30}$&$\rm Log\, \nu_{\rm sync}\sim \mathcal{U}(12, 16)$\\ \vspace{0.05cm}
                        & $\alpha_{\rm sync}$ & -- & $-0.61\pm 0.01$&$\alpha_{\rm sync}\sim\mathcal{U}(-0.9, -0.4)$\\ \vspace{0.35cm}
                        & Log [$F_{\rm sync}$ / 1~mJy] & -- & $-1.11_{-0.27}^{+0.18}$&$\rm Log \, F_{\rm sync} \sim \mathcal{U}(-3, 2)$\\  \vspace{0.05cm}
Discrete ejecta  & $\alpha_{\rm thin}$  & $-0.78_{-0.43}^{+0.24}$ & --&$\alpha_{\rm thin}\sim \mathcal{U}(-1.5, 0)$\\ \vspace{0.35cm}
                & Log [$F_{\rm thin\, ,10}$ / 1~mJy]  & $-1.03\pm 0.05$ & --&$\rm Log \, F_{\rm thin\, ,10} \sim \mathcal{U}(-5, 2)$\\  \vspace{0.05cm}
                & Bayesian p-value  & 0.07 & 0.78 &--\\ \hline
    \end{tabular}
    \label{Tab:fit_res}\\
    \parbox{1.0\textwidth}{
    {\bf Notes.} The median values of the posterior distributions of the model parameters are reported, along with lower and upper uncertainties. These uncertainties correspond to the 15.9th and 84.1th percentiles of the posterior distributions for each parameter. The last column reports the prior bounds that were used for all parameters; $\mathcal{N}(\mu, \sigma)$ is a normal distribution centred in $\mu$ and with variance $\sigma^2$, and $\mathcal{U}(a,b)$ is a uniform distribution in the $(a,b)$ interval.
    $^{\ast}$ Value from \cite{shahbaz22}.
$^{\dagger}$ Value from \cite{bogdanov18}.
}
\end{table*}
%%%%%%%%%%%%%%%%%%%%%%%%%%%%%%%%%%%%%%%%%%%%%%%%%%%%%%%%%%%%%%%%%%%%%%%%

%%%%%%%%%%%%%%%%%%%%%%%%%%%%%%%%%%%%%%%%%%%%%%%%%%%%%%%%%%%%%%%%%%%%%%%%
\begin{figure*}
    \centering
    \includegraphics[width=1.0\textwidth]{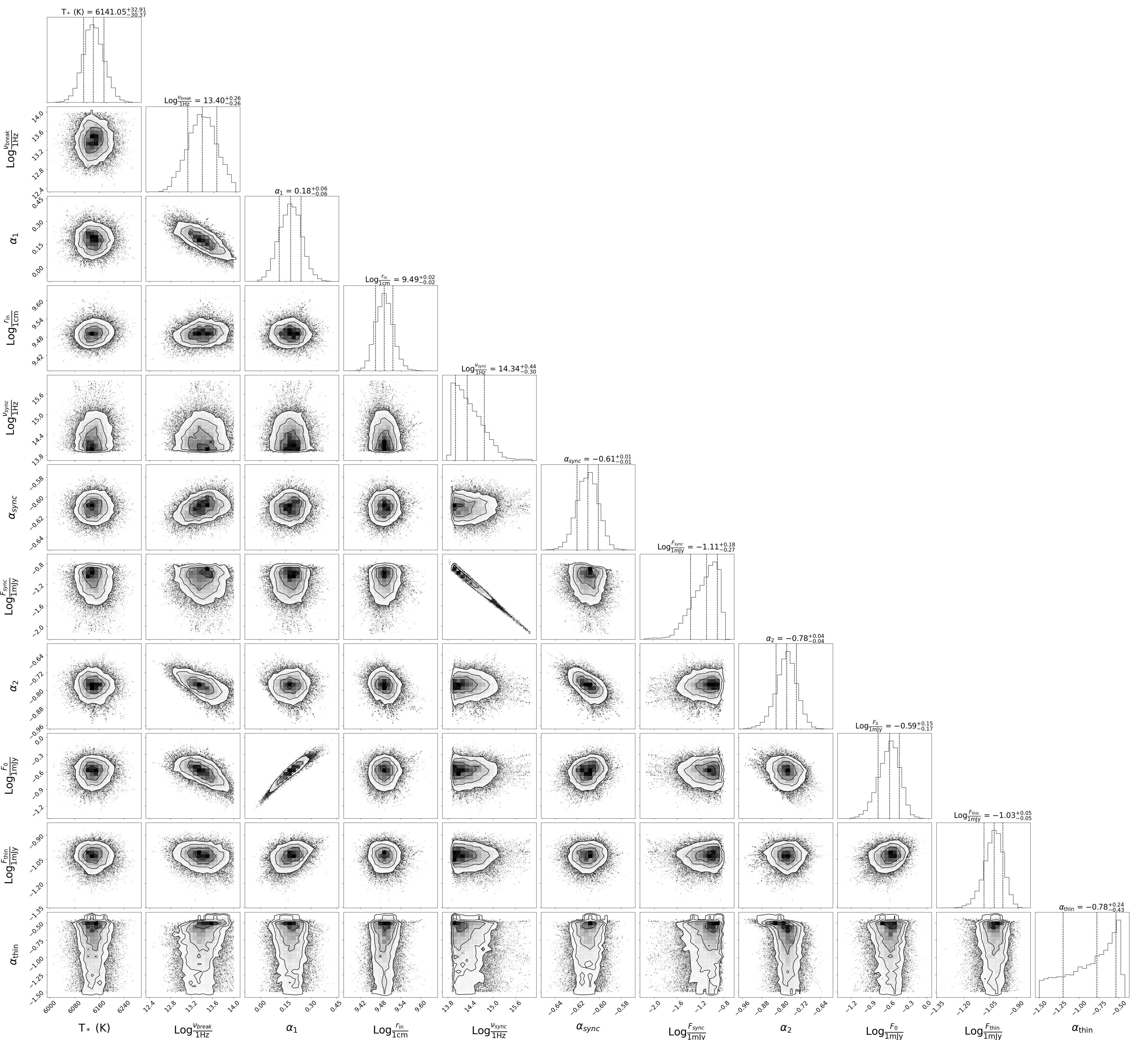}
    \vspace{0.05cm}
    \caption{Corner plot showing the posterior probability distributions of the parameters obtained from the MCMC sampling algorithm. The solid lines in the panels along the diagonal represent the probability density of each parameter. The vertical dashed lines indicate the 16th, 50th, and 84th percentiles of the distributions. The median of the parameter posterior and the 1$\sigma$ error bars are displayed above each panel. The off-diagonal panels show the 2D posterior distributions, with contour lines indicating the 1$\sigma$, 2$\sigma$, and 3$\sigma$ equivalent bounds.}
    \label{fig:cornerplot}
\end{figure*}
%%%%%%%%%%%%%%%%%%%%%%%%%%%%%%%%%%%%%%%%%%%%%%%%%%%%%%%%%%%%%%%%%%%%%%%%

\textit{High mode.} We used a more complicated model following the physical picture explained above. In particular, we considered that the X-rays are produced by the sum of the optically thin component of the synchrotron emission from the base of the compact jet and the synchrotron emission from the region of shock between the pulsar wind and the inner flow. For the latter, we used the analytical expression for the synchrotron spectrum by \cite{chevalier98} (see his Eq.\,1). 
For the emission at lower frequencies, we included the blackbody of the irradiated star (fixing $L_{\rm irr}$ to $6.5\times10^{33}$\,\lum), the multi-colour blackbody of the accretion disc, and the optically thick component of the synchrotron emission from the compact jet. The parameters of the fit are $T_{*}$, $\nu_{\rm break}$, $\alpha_1$, $\alpha_2$, $F_0$, $r_{\rm in,opt/UV}$, $\nu_{\rm sync}$, $\alpha_{\rm sync}$ and $F_{\rm sync}$, where $\nu_{\rm sync}$, $\alpha_{\rm sync}$, $F_{\rm sync}$ are the peak frequency of the synchrotron emission, the slope of the optically thin synchrotron emission and the normalisation of the synchrotron emission, respectively.

\subsubsection{Model fit}
Employing both datasets (high and low mode), we performed an MCMC sampling of the posterior probability density function for the parameter space of the two models simultaneously (since some of the free parameters are shared by both models).
We used a Gaussian prior for the temperature of the star centred on the value measured by \cite{shahbaz22} during the disc state, $T_*$ = ($6128\pm33$)\,K. 
Gaussian priors have also been used for the parameters $\alpha_1$ (centred on $0.2\pm0.2$) and $\alpha_2$ (centred on $-0.7\pm0.2$), following the results of \cite{bogdanov18}. 
We note that we also tried wider Gaussian priors for these parameters, resulting in wider posterior distributions with medians compatible with the results below within 1--2$\sigma$.
For all other parameters, we used uniform priors in very wide intervals to sample a large region of the parameter space (see the last column of Table\,\ref{Tab:fit_res}). This is of great importance to minimise the probability that the actual source parameters fall outside the region sampled by the chosen priors. 

\subsubsection{Results}
The results of the fit are reported in Table\,\ref{Tab:fit_res} and are summarised in the following paragraphs. All the parameters have been estimated as the median of the marginal posterior distributions, with $1\sigma$ credible intervals coming from the 16th--84th percentiles of the posterior distributions (see Fig.\,\ref{fig:cornerplot} for the histograms of the parameter samples).
%The temperature of the irradiated companion star is in line with that reported in previous works and that estimated in the radio pulsar state (e.g. \cite{cotizelati14,shahbaz22}). The power-law slopes below and above the jet break frequency are consistent with optically thick and thin synchrotron emission from a compact jet, respectively. In particular, the slope of the optically thick component is comparable with those reported by \cite{bogdanov18} within the uncertainties. 
Most of the marginal posterior distributions of the parameters shown in Fig.\,\ref{fig:cornerplot} are nearly Gaussian, meaning that they are almost symmetrical.
Figure\,\ref{fig:cornerplot} also shows a strong correlation between $\nu_{\rm sync}$ and $F_{\rm sync}$, and between $F_0$ and $\alpha_1$. This is however not surprising considering the scarce number of points at the lowest frequencies and the impossibility to constrain quantities such as $\nu_{\rm break}$ and $\nu_{\rm sync}$, due to numerous overlapping components at infrared and optical frequencies. In particular, we note that $F_{\rm sync}$, $\nu_{\rm sync}$ and $\alpha_{\rm thin}$ have a non-symmetrical posterior distributions, with long tails towards the upper or the lower limit of the chosen interval for the prior, depending on the parameter. However, the peak of their distributions is well determined (see Fig.\,\ref{fig:cornerplot}).  
To estimate the goodness of the fit, we used the Bayesian p-value (reported in the last line of  Table\,\ref{Tab:fit_res}; \citealt{Lucy2016}). The p-value is 0.78 for the high mode and 0.07 for the low mode, meaning that the results of the fit are indeed acceptable (with a slight indication of overestimated and underestimated uncertainties for the high and low mode fit, respectively).

The synchrotron emission caused by the shock has a higher flux density in the optical band than the average pulsed flux density fraction of $\sim$1\% estimated by \cite{papitto19}, which makes our scenario plausible. In particular, we find that the shock contributes $\sim$3\% of the emission in the optical band in the high mode, which is similar to the contribution of the optically thin component of the compact jet emission. In the \xmm/EPIC 0.3--10\,keV range, the shock accounts for $\sim$83\% of the emission, while the jet only contributes $\sim$17\% of the flux in the high mode.

\subsection{Optical polarisation measurements and predictions of X-ray polarisation}
The average optical polarisation reported in this work is similar to levels previously measured for this source \citep{baglio16,hakala18}. Since the polarisation level is $<$1\%, it could have multiple origins. Given the lack of modulation of the Stokes parameters with the orbital period, it is unlikely that the source of the polarisation is scattering of radiation in any region of the system (such as the surface of the accretion disc), in contrast to previous suggestions \citep{baglio16}.
We identify two possible main emission mechanisms for the polarised emission, both of which are non-thermal: optically thin synchrotron emission from the compact jet, or synchrotron emission at the shock between the pulsar wind and the accretion inflow, which also gives rise to the observed optical pulsations (as also suggested by \citealt{hakala18}).
The jet flux density in the $r$-band is $\sim$0.02\,mJy, corresponding to a fractional contribution to the total flux in the same band of $\sim$2.4\% and $\sim$2\% in the high and low modes, respectively.
In the high mode, where the polarisation level measured in the $R$-band is 0.65$\pm$0.43\%, this translates into a polarisation level of the jet emission of $\sim$26\% (or $\sim$37\% if the average polarisation level of 0.92\% is considered instead). In the low mode, the jet emission is instead intrinsically polarised at a level of $\sim$17\%.
Following \cite{Shahbaz2019book} and references therein, we estimate a level of ordering of the magnetic field lines of $\sim$30\% (considering the case of an optically thin synchrotron emission with a spectral slope of $\sim$-0.78, as obtained from our fit and averaging across modes; see Table\,\ref{Tab:fit_res}). This indeed indicates a highly ordered magnetic field and suggests that tMSPs might have a more ordered magnetic field at the base of the jet than other NS and black hole XRBs \citep{russell08,baglio14,russell16,baglio20}. Based on our picture, this could be ascribed to the presence of a continuous, steady flow of plasma near the pulsar, which may help maintain a more ordered magnetic field structure at the base of the jet.
%In the first case, the very low level of average optical polarisation ($0.92 \pm 0.39\%$) would imply a strongly tangled magnetic field at the base of the jet, as it is often observed in neutron star X-ray binaries (e.g. \cite{baglio20,baglio14}).
%In addition, according to our SED modeling (see Fig.\,\ref{fig:SED}), the contribution of the optically thin synchrotron radiation from the jet to the total flux in the X-rays is $100\%$ in the low mode and $\approx$20\% in the high mode.
%at 5.15\,keV (mid point of the NICER energy band, 0.3--10\,keV). 
Assuming that the observed optical polarisation is due to this emission component and that the same population of accelerated electrons in the jet produces optically thin synchrotron radiation at all frequencies above the break frequency, we can predict a polarisation level in the X-ray band -- in the 2--8\,keV energy range of the Imaging X-ray Polarimetry Explorer (IXPE; \citealt{Weisskopf2016,soffitta21}) due to the jet of $\approx$17\% in the low mode, where the jet contributes to $\sim$100\% of the flux according to our SED modelling. 
%%This would imply a strongly tangled magnetic field at the base of the jet, as already inferred from the measured optical LP.

Alternatively, the observed optical polarisation might originate from synchrotron radiation at the region of shock between the pulsar wind and the accretion flow. In this scenario, polarisation should only be detected during the high mode. Unfortunately, the significance of our polarisation results obtained in the high and low modes separately is not high enough to verify this, as the results are compatible with each other within $1\sigma$ ($0.36^{+0.27}_{-0.35}\%$ in the low mode and 0.65$\pm$0.43\% in the high mode). 
The fractional flux contribution of the shock in the $r$-band according to the SED modelling in the high mode is $\sim$4.8\%, which translates into a LP of the emission of the shock of $\sim$15--20\%, with peaks of up to 65--90\% at the pulsations (in fact, the fraction of pulsed radiation is typically of $\sim$1\%; \citealt{papitto19}). This prediction could be tested using next-generation optical high-time resolution polarimetric instruments  \citep{Ghedina2022}.
In the X-rays instead, we predict a fractional flux contribution of the shock emission of $\sim$90\% in the IXPE energy range, which converts into an intrinsic LP level of $\approx$12--17\%. This could be detected in an IXPE observation of $\sim$460\,ks (assuming \src\ spends 70\% of the time in the high mode).

In both scenarios, the validity of these predictions depends on the assumption that the optical polarisation is solely caused by synchrotron radiation from the jet or shock. However, if the observed polarisation is due to a combination of these factors or if other processes are at play that are not being considered (such as scattering on the disc surface contributing to the modulation with a low S/N in our measurements), these predictions may need to be revised.

\subsection{The millimetre continuum emission of \src\ and other X-ray binary systems}
We have reported the detection of millimetre continuum emission for the first time from \src\ (and from a tMSP in general) using ALMA observations at $\approx$100\,GHz. The millimetre emission is unresolved, implying a size for the emission region smaller than $\approx6\times10^{15}$~cm (this corresponds to $\approx7\times10^{4}$ times the orbital separation of the system; \citealt{archibald13}). 
Millimetre/sub-millimetre continuum emission has been firmly detected from a number of transiently accreting stellar-mass black holes \citep{paredes00b,ogley00,fender00,fender01,ueda02,russell13,vanderhorst13,tetarenko15,tetarenko17,tetarenko19,russell20,dehaas21,koljonen21,tetarenko21,diaztrigo21} and in three accreting NSs so far: the persistent systems 4U~1728$-$34 and 4U~1820$-$30 \citep{diaztrigo17} and the transient accreting millisecond pulsar Aql X-1 \citep{diaztrigo18}. In these cases, however, such emission has been detected during accretion states that are seemingly different from the X-ray sub-luminous disc state attained by \src\ at the epoch of our ALMA observations, as detailed below. 

In black hole systems, millimetre continuum emission is typically detected during the rising phase of an X-ray outburst or along the outburst decay, during the `hard' X-ray spectral state. In this state, the X-ray emission is typically dominated by a power-law component that is believed to arise from Compton up-scattering of soft disc photons by a population of hot thermal electrons located either in the innermost region of the accretion flow or in an extended cloud above the disc (for a recent review, see \citealt{motta21}).
At low frequencies, a flat to slightly inverted optically thick spectrum ($\alpha>0$) has been observed from radio through sub-millimetre frequencies and above and has been associated with partially self-absorbed synchrotron radiation from electron populations in different regions along a compact jet. 
This spectrum becomes optically thin ($\alpha<0$) typically around infrared frequencies, resulting in a spectral break \citep{russell13a,russell13b}. On the other hand, at very high mass accretion rates (typically close to the outburst peak), these systems can be in a `soft' X-ray spectral state where the X-ray emission is dominated by a thermal component produced in the hot inner regions of the accretion disc (see \citealt{motta21} and references therein). In the state that immediately precedes or follows the soft state during the outburst (the so-called soft-intermediate state), the optically thick jet emission is observed to be quenched altogether, while discrete jet ejecta characterised by an optically thin steep spectrum ($\alpha<0$) in the radio band can be launched \citep{fender04,russell20} and be accompanied by strong radio variability and bright multi-frequency flaring activity (even in the sub-millimetre band; \citealt{tetarenko19}).

%%%%%%%%%%%%%%%%%%%%%%%%%%%%%%%%%%%%%%%%%%%%%%%%%%%%%%%%%%%%%%%%%%%%%%%%
\begin{table*}
%\footnotesize
\caption{
\label{tab:mmxray}
Millimetre and X-ray luminosities of accreting compact objects in XRBs.}
\centering
\begin{tabular}{ccccc}
\hline\hline
Source              & Distance    & Millimetre luminosity$^{\rm a}$           & X-ray luminosity$^{\rm b}$            & Reference \\
                            & (kpc)           & (erg\,s$^{-1}$)                                     & (erg\,s$^{-1}$)                               &  \\
\hline
MAXI\,J1659$-$152  & 6                & (7$\pm$1)$\times$10$^{31}$        & (1.389$\pm$0.001)$\times$10$^{37}$    & \cite{vanderhorst13} \\
MAXI\,J1659$-$152  & 6                & (5$\pm$1)$\times$10$^{31}$        & (4.60$\pm$0.01)$\times$10$^{37}$      & \cite{vanderhorst13} \\
MAXI\,J1659$-$152  & 6                & $<$2.6$\times$10$^{31}$           & (4.81$\pm$0.01)$\times$10$^{37}$      & \cite{vanderhorst13} \\
MAXI\,J1659$-$152  & 6                & $<$2.7$\times$10$^{31}$           & (4.67$\pm$0.02)$\times$10$^{37}$      & \cite{vanderhorst13} \\
V404 Cygni         & 2.39             & (3.64$\pm$0.06)$\times$10$^{30}$  & (8$\pm$2)$\times$10$^{34}$            & \cite{tetarenko19} \\
V404 Cygni         & 2.39             & (9.4$\pm$0.6)$\times$10$^{29}$    & (1.5$\pm$0.1)$\times$10$^{34}$        & \cite{tetarenko19} \\
MAXI\,J1535$-$571  & 4.1              & (4.7$\pm$0.2)$\times$10$^{32}$    & (2.00$\pm$0.01)$\times$10$^{38}$      & \cite{russell20} \\
MAXI\,J1535$-$571  & 4.1              & (1.0$\pm$0.2)$\times$10$^{32}$    & (8.6$\pm$0.1)$\times$10$^{38}$        & \cite{russell20} \\
GX\,339$-$4        & 8                & (1.53$\pm$0.08)$\times$10$^{31}$  & (1.35$\pm$0.08)$\times$10$^{36}$      & \cite{dehaas21} \\ 
%MAXI\,J1820$+$070  & 2.96            & $\pm$             & $\pm$             & \cite{tetarenko18}  \\
MAXI\,J1820$+$070  & 2.96             & (9.651$\pm$0.004)$\times$10$^{31}$ & (3.50$\pm$0.01)$\times$10$^{37}$     & \cite{tetarenko21}  \\
GRS\,1915$+$105    & 8.6              & 2.2$\times$10$^{31}$             & (2.07$\pm$0.03)$\times$10$^{37}$      & \cite{koljonen21}  \\
GRS\,1915$+$105    & 8.6              & 2.6$\times$10$^{31}$             & (2.69$\pm$0.01)$\times$10$^{37}$       & \cite{koljonen21}  \\
XTE\,J1118$+$480   & 1.7          & 6.9$\times$10$^{30}$              & 2.4$\times$10$^{35}$                  & \cite{fender01}   \\ 
\hline
4U\,1820$-$30$^{\rm c}$      & 6.4            & (1.67$\pm$0.09)$\times$10$^{30}$    & (3.22$\pm$0.03)$\times$10$^{37}$    & \cite{diaztrigo17} \\
Aql X-1                    & 5.2              & (2.06$\pm$0.08)$\times$10$^{30}$    & (1.73$\pm$0.01)$\times$10$^{37}$    & \cite{diaztrigo18} \\
Aql X-1                    & 5.2              & (1.31$\pm$0.08)$\times$10$^{30}$    & (9.2$\pm$0.1)$\times$10$^{37}$      & \cite{diaztrigo18} \\
Aql X-1                    & 5.2              & (2.44$\pm$0.09)$\times$10$^{30}$    & (6.65$\pm$0.07)$\times$10$^{36}$    & \cite{diaztrigo18} \\
\srcfirst\ (mean)  & 1.37             & (3.2$\pm$0.2)$\times$10$^{28}$      & (1.7$\pm$0.2)$\times$10$^{33}$      & This work \\
\srcfirst\ (high)  & 1.37             & (2.0$\pm$0.3)$\times$10$^{28}$      & (2.40$\pm$0.04)$\times$10$^{33}$      & This work \\
\srcfirst\ (low)   & 1.37             & (3.0$\pm$0.5)$\times$10$^{28}$      & (3.6$\pm$0.4)$\times$10$^{32}$      & This work \\
\hline    
\end{tabular}\\
\parbox{1.0\textwidth}{
{\bf Notes.} Values are reported for those systems that have been clearly detected in the millimetre/sub-millimetre bands, have a known distance, and have simultaneous or nearly simultaneous millimetre/sub-millimetre and X-ray observations. Systems above the horizontal line are black hole systems, while those below the line are NS systems.
$^{\rm a}$ Millimetre luminosities are evaluated at a frequency of 100\,GHz. A flat spectrum was assumed to estimate the luminosity for those cases where the flux density is available at other frequencies in the literature, except for MAXI\,J1820$+$070 ($\alpha$=0.25) and 4U\,1820$-$30 ($\alpha$=0.13).
$^{\rm b}$ X-ray luminosities are evaluated over the 1--10\,keV energy range and are derived from \swift/XRT observations except for the case of \src, for which values are taken from Table\,\ref{tab:mode_properties}. Spectra were extracted using the \swift/XRT data products generator \citep{evans09} and fitted using an absorbed power-law model. For the second pair of simultaneous observations of GRS\,1915$+$105, the X-ray luminosity is evaluated using NICER Obs. ID 2596010301.
$^{\rm c}$ The millimetre and X-ray observations were separated by 4 days.
}
\end{table*}
%%%%%%%%%%%%%%%%%%%%%%%%%%%%%%%%%%%%%%%%%%%%%%%%%%%%%%%%%%%%%%%%%%%%%%%%

%%%%%%%%%%%%%%%%%%%%%%%%%%%%%%%%%%%%%%%%%%%%%%%%%%%%%
\begin{figure*}[!h]
\begin{center}
\includegraphics[width=1.0\textwidth]{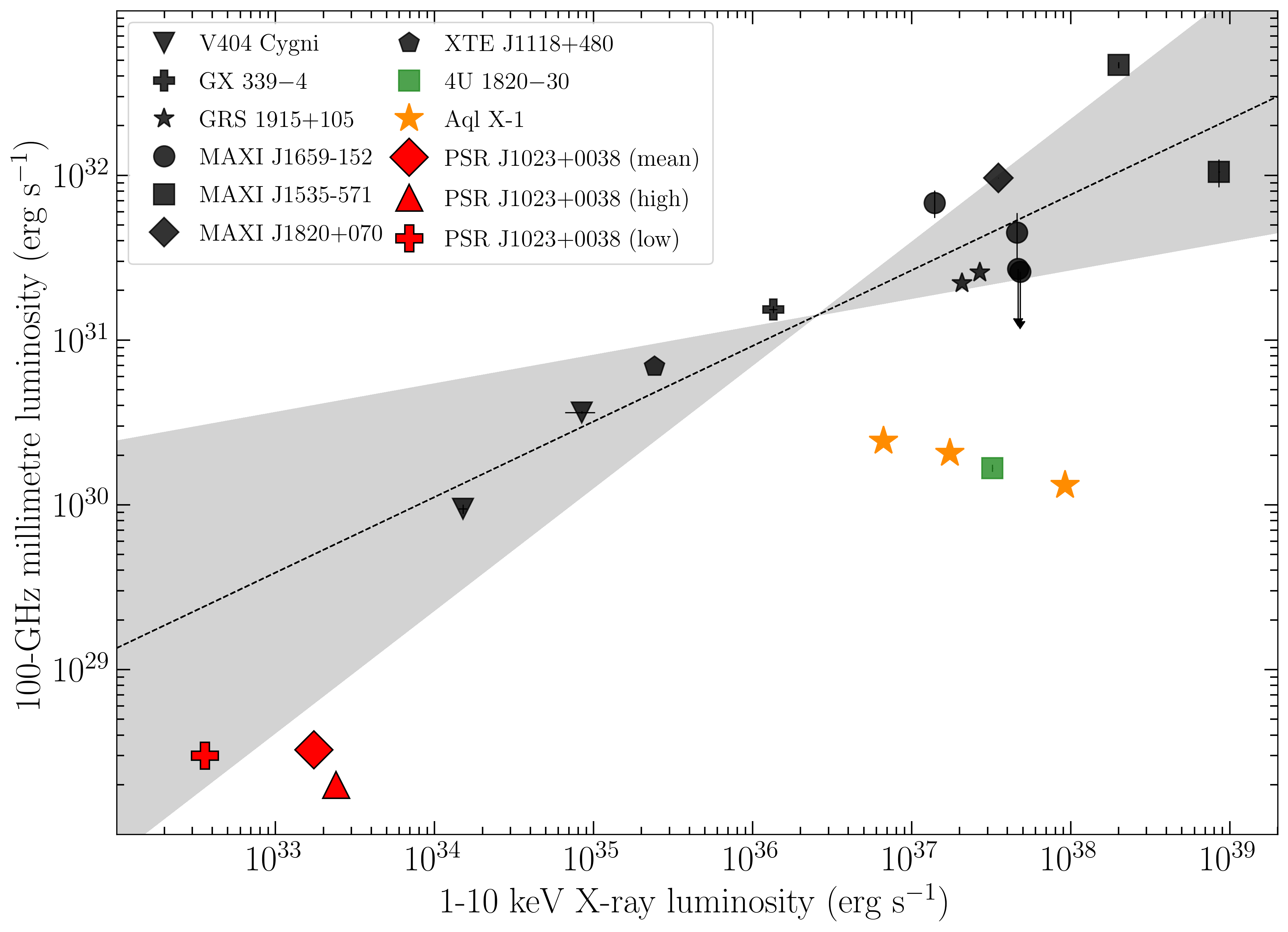}
\caption{Millimetre and X-ray luminosities for accreting compact objects in XRBs. Black hole systems are marked in black. The adopted values for the luminosities are listed in Table\,\ref{tab:mmxray}. The dashed black line and the grey shaded area show the best-fitting slope for the black hole correlation and the region on the plane at less than 3$\sigma$ from the best-fitting slope, respectively (see the main text for details).
} 
\label{fig:mmxray}
\end{center}
\end{figure*}
%%%%%%%%%%%%%%%%%%%%%%%%%%%%%%%%%%%%%%%%%%%%%%%%%%%%%

The picture for NS systems is more diverse (see \citealt{vandene21} for a recent overview). All three of the above-mentioned NS systems were observed in the millimetre band with ALMA, along with observations at other frequencies, in regimes of high mass accretion rates ($L_X>5\times10^{36}$\,\lum). 4U~1728$-$34 was observed during a transition from a hard to a soft state. The low-frequency spectrum was compatible with synchrotron emission from a compact jet and also revealed the presence of a break from optically thick to optically thin synchrotron radiation at frequencies in the range 10$^{13}$ -- 10$^{14}$\,Hz. In the case of 4U~1820$-$30, enhanced millimetre emission was detected during a soft X-ray state (the first time millimetre emission has been observed for an accreting NS in the soft state). The spectrum from the radio to the sub-millimetre bands was slightly inverted and showed no evidence for the presence of a spectral break \citep{diaztrigo17}. This indicates emission from a compact jet and supports earlier findings that jets in NS systems do not all appear to be completely suppressed in soft accretion states \citep{migliari04}. Aql X-1 was observed five times using ALMA over a time span of $\sim$1.5 months during the decay phase of an X-ray outburst in 2016, covering multiple transitions between distinct accretion states. The spectrum from the radio to the millimetre bands was again compatible with emission from a compact jet and, for the first time in an accreting NS in a low-mass XRB, a break in the SED was detected at a frequency that varies dramatically depending on the accretion state. Specifically, the break frequency initially decreased from $\sim$100\,GHz to less than $\sim$5\,GHz during a transition from a hard to a soft state, and then increased again up to a frequency within the range $\sim$30--100\,GHz during a transition to a hard state at later stages of the outburst decay \citep{diaztrigo18}. 

Unlike all the cases mentioned above, the ALMA observations of \src\ were performed at an epoch when the system did not display any changes in its spectral state and was lingering in a sub-luminous X-ray state ($\approx$10$^{33}$\,\lum). It is worth noting that both the average and the mode-resolved X-ray spectra of \src\ have always been hard and have not shown any substantial changes over the past ten years or so during the sub-luminous X-ray state.

We can compare the millimetre and X-ray luminosities of \src\ with those measured for black-hole and NS systems using quasi-simultaneous millimetre and X-ray observations. These luminosities are reported in Table\,\ref{tab:mmxray} and plotted in Fig.\,\ref{fig:mmxray}. Despite the relatively low amount of existing millimetre/sub-millimetre data for XRBs, Fig.\,\ref{fig:mmxray} shows that millimetre continuum emission has been detected in systems that attain a wide range of X-ray luminosities ($\sim$10$^{33}$--10$^{39}$\,\lum) and that \src\ currently holds the record as the faintest XRB firmly detected in the millimetre band.

%The bulk of the mm emission detected from \src\ is expected to arise from the compact jet in both modes, since the optically thin emission from the discrete ejecta in the low mode is negligible at these frequencies.
%see bottom panel of Fig.\,\ref{fig:SED}), 
The mechanism responsible for the X-ray emission in the two modes in \src\ is clearly different from that at work in black hole systems in the hard state, so a proper comparison cannot be made. With this in mind, we assessed how the position of \src\ on the $L_{\rm mm}$ -- $L_{\rm x}$ plane compares with those of black hole systems by parameterised the dependence of $L_{\rm mm}$ upon $L_{\rm x}$ as a relation of the form ${\rm Log}L_{\rm mm} - {\rm Log}L_{\rm mm,ave} = \alpha + \beta({\rm Log}L_{\rm X} - {\rm Log}L_{\rm X,ave})$\footnote{$L_{\rm mm,ave}=10^{31.93}$\,\lum\ and $L_{\rm X,ave}=10^{38.03}$\,\lum\ are the average values for the detections in our black hole sample.}. We then investigated the shape of the $L_{\rm mm}$ -- $L_{\rm x}$ correlation for black hole systems by performing a Bayesian-based MCMC sampling of the black hole data (including upper limits) using the linear regression algorithm \textsc{linmix}$_-$\textsc{err} \citep{kelly07}. There are three free parameters: the intercept $\alpha$, the slope $\beta$ and $\sigma_0$, a parameter that accounts for an intrinsic random (Gaussian) scatter of the values around the best-fit model. We calculated the median values of the fit parameters from 10$^4$ draws from the posterior distribution, and determined the 1$\sigma$ credible intervals of these parameters using the 16th--84th percentiles of the posterior distributions. We derived $\alpha= -0.04\pm0.15$, $\beta= 0.46\pm0.09$ and $\sigma_0=0.18\pm0.13$. The best-fitting slope is consistent with that derived for the $L_{\rm radio}$ -- $L_{\rm x}$ correlation for black hole systems \citep{gallo18} within the uncertainties.
Figure\,\ref{fig:mmxray} shows the best-fitting slope (black dashed line) and the region on the plane where the values of $L_{\rm mm}$ and $L_{\rm X}$ are located at a distance of less than 3$\sigma$ from the best-fitting slope (grey shaded area). This figure shows that, during the low mode, \src\ follows the same correlation as black hole systems in the hard state \citep{bogdanov18}.

\section{Summary and conclusions}
\label{sec:conclusions}
We have presented the results of the largest multi-wavelength campaign ever conducted on the tMSP \srcfirst. This campaign took place in June 2021 while the system was in an active low-luminosity X-ray disc state. Our key findings can be summarised as follows:

$\bullet$ On the first night of observations, flaring activity was detected at X-ray and optical wavelengths, followed by mode-switching behaviour in X-rays. The UV emission was significantly fainter in low mode, while the radio continuum emission was brighter. No evidence of pulsed radio emission was detected in either mode. Our observations provide the most stringent limits on the pulsed radio flux density to date.

$\bullet$ On the second night of observations, a low-significance variation in the optical polarised signal was detected during mode changes, with a slightly higher LP level in the high mode compared to the low mode. The millimetre emission did not strictly follow the X-ray low modes, except for two brief flares. One of these flares occurred at the high-to-low mode switch, while the exact onset time of the other flare could not be determined due to a lack of observational coverage. 

$\bullet$ These results allow us to draw a physical picture of high-low mode switches in \src, which involves a rotation-powered pulsar, an accretion disc, and discrete mass ejections on top of a relativistic compact jet. During the high mode, synchrotron radiation at the shock front between the pulsar wind and the inner accretion flow produces most of the X-ray emission, as well as X-ray, UV, and optical pulsations. The switch to low mode is triggered by discrete mass ejections, resulting in a drop in the X-ray flux and a quenching of pulsations. During the low mode, the pulsar wind still penetrates the accretion disc and launches a compact jet, contributing to the multi-band emission in the same way as in the high mode. Eventually, the flow from the disc refills the inner regions outside the NS light cylinder, leading to increased emission and pulsations at X-ray, UV, and optical frequencies. The system then switches back to the high mode. This scenario is supported by the modelling of the SEDs extracted separately during the high and low modes.

$\bullet$ Relativistic pulsar winds can shape the structure and dynamics of accretion flows in XRB systems that contain quickly spinning NSs and can be used to probe emission mechanisms in tMSPs and other XRB systems.

$\bullet$ Black hole systems and tMSPs  share intriguing similarities in their phenomenological properties, which emphasises the need for further research to deepen our understanding of accretion physics in compact objects.

\noindent These results show how multi-wavelength campaigns that combine diverse observational techniques provide powerful tools for unravelling the nature of elusive objects such as tMSPs.
 
\begin{acknowledgements}
We thank the referee for helpful comments.
We thank the ALMA staff, particularly Edwige Chapillon, for the help with scheduling the observations; the NICER PI, Keith Gendreau, for approving our ToO request and the operation team for executing the observations; Brad Cenko and the \swift\ duty scientists and science planners, for making the \swift\ Target of Opportunity observation possible; Nando Patat, for the help with scheduling the NTT/SOFI observations; Giulia Illiano, for providing checks on timing analysis of X-ray data; Rob Fender, for providing useful notebooks on synchrotron radiation from variable radio sources; Nanda Rea, for helpful comments on the manuscript.\\

The results reported in this study are based on observations obtained with XMM-Newton (PI: Campana); NuSTAR (PI: Campana); the NASA/ESA Hubble Space Telescope (obtained from the Space Telescope Science Institute, which is operated by the Association of Universities for Research in Astronomy, Inc., under NASA contract NAS 5–26555. These observations are associated with program 16061; PI: Campana); FAST (PI: Hou); the REM Telescope, INAF Chile (PI: Baglio).  Part of the observations presented in this study were obtained at the European Southern Observatory under ESO programme 107.22RK.001 (PI: Baglio). This paper makes also use of the following ALMA data: ADS/JAO.ALMA\#2019.A.00036.S (PI: Coti Zelati). 
\xmm\ is an European Space Agency (ESA) science mission with instruments and contributions directly funded by ESA member states and NASA.
\nustar\ is a project led by the California Institute of Technology, managed by the Jet Propulsion Laboratory and funded by NASA.
The National Radio Astronomy Observatory is a facility of the National Science Foundation operated under cooperative agreement by Associated Universities, Inc.
FAST is a Chinese national mega-science facility, operated by National Astronomical Observatories of the Chinese Academy of Sciences (NAOC).
NICER is a 0.2--12\,keV X-ray telescope operating on the International Space Station, funded by NASA. 
The Neil Gehrels Swift Observatory is a NASA/UK/ASI mission.
ALMA is a partnership of ESO (representing its member states), NSF (USA) and NINS (Japan), together with NRC (Canada), MOST and ASIAA (Taiwan), and KASI (Republic of Korea), in cooperation with the Republic of Chile. The Joint ALMA Observatory is operated by ESO, AUI/NRAO and NAOJ. \\

The XMM-Newton SAS is developed and maintained by the Science Operations Centre at the European Space Astronomy
Centre. The NuSTAR Data Analysis Software (NuSTARDAS) is jointly developed by the ASI Science Data Center (ASDC, Italy) and the California Institute of Technology (Caltech, USA).\\

The data that support the findings of this study are publicly available at their respective online archive repositories (\xmm: \url{http://nxsa.esac.esa.int/nxsa-web/}; NICER: \url{https://heasarc.gsfc.nasa.gov/FTP/nicer/data/obs}; \nustar: \url{https://heasarc.gsfc.nasa.gov/W3Browse/all/numaster.html}; \swift: \url{https://heasarc.gsfc.nasa.gov/cgi-bin/W3Browse/w3browse.pl}; HST: \url{http://hst.esac.esa.int/ehst/}; VLA: \url{https://data.nrao.edu/portal/}; VLT/FORS2: \url{http://archive.eso.org/scienceportal/}; ALMA: \url{http://almascience.nrao.edu/aq/}; REM: \url{http://ross.oas.inaf.it/REMDB/public.php}; NTT/SOFI: \url{http://archive.eso.org/scienceportal/}). FAST data can be accessed through the FAST Data Center: \url{https://fast.bao.ac.cn/cms/article/11/}.\\

FCZ is supported by a Ram\'on y Cajal fellowship (grant agreement RYC2021-030888-I) and Catalan grant SGR-Cat 2021 (PI: Graber).
FCZ and DFT are supported by the program Unidad de Excelencia María de Maeztu CEX2020-001058-M.
FCZ, SCa, PD'A, FA, PC, DdM and AP acknowledge financial support from INAF-Fundamental research astrophysics project ``Uncovering the optical beat of the fastest magnetised neutron stars'' (FANS; PI: AP).
SCa, SCo and PD'A acknowledge support from ASI grant I/004/11/5.
GB acknowledges support from the PID2020-117710GB-I00 grant funded by MCIN/ AEI /10.13039/501100011033.
AMZ is supported by PRIN-MIUR 2017 UnIAM (Unifying Isolated and Accreting Magnetars, PI S. Mereghetti).
XH is supported by the National Natural Science Foundation of China through grant No. 12041303.
JL is supported by the National Natural Science Foundation of China through grant No.12273038.
DdM and AP acknowledge financial support from the Italian Space Agency (ASI) and National Institute for
Astrophysics (INAF) under agreements ASI-INAF I/037/12/0 and ASI-INAF n.2017- 14-H.0, from INAF `Sostegno alla ricerca scientifica main streams dell`INAF', Presidential Decree 43/2018 and from INAF `SKA/CTA projects', Presidential Decree 70/2016.
DMR and DMB acknowledge the support of the NYU Abu Dhabi Research Enhancement Fund under grant RE124.
DFT is supported by the grants PID2021-124581OB-I00 funded by MCIN/AEI/10.13039/501100011033, 2021SGR00426 of the Generalitat de Catalunya and by 
MCIN with funding from European Union NextGeneration EU (PRTR-C17.I1).
MM acknowledges the research programme Athena with project number 184.034.002, which is (partly) financed by the Dutch Research Council (NWO) and thanks the Team Meeting at the International Space Science Institute (Bern) for fruitful discussions. 
FMV acknowledges support from the Spanish Ministry of Science for the grant PID2020–120323GB–I00 and for the  FJC2020-043334-I grant financed by MCIN/AEI/10.13039/501100011033 and Next Generation EU/PRTR. 
This work was also partially supported by the COST Action `PHAROS' (CA 16124).\\

Software:
APLPY v.2.1.0 \citep{robitaille12};
Astropy v.5.3 \citep{astropy:2013, astropy:2018, astropy:2022}; 
CASA v.6.5.2 \citep{casa}; 
CIAO v.4.15 and CALDB v.4.10.4 \citep{fruscione06};
CORNER PLOT v.2.2.1 \citep{corner};
DAOPHOT \citep{Stetson1987}, part of the Starlink software \citep{currie14,berry22};
HEASOFT v.6.31.1 \citep{heasoft14}; 
IRAF v.2.17 (\url{https://github.com/iraf-community/iraf});
LINMIX$_-$ERR (\url{https://github.com/jmeyers314/linmix});
MATPLOTLIB v.3.7.1 \citep{hunter07};
NICERDAS v.10a;
NUMPY v.1.24.0 \citep{harris20}; 
NUSTARDAS, v.2.1.2;
PRESTO v.4.0 (\url{https://github.com/scottransom/presto});
SAOImageDS9 v.8.4 \citep{joye03};
SAS v.20.0 \citep{gabriel04};
SCIPY v.1.10.1 \citep{scipy20}; 
SRPAstro v.4.8.0 (\url{https://pypi.org/project/SRPAstro/});
XRONOS v.5.22 \citep{stella92}; 
XSPEC v.12.13.0g \citep{arnaud96}.\\
Upon request, the corresponding authors (MCB and FCZ) will provide the code used to produce the figures.
\end{acknowledgements}

\bibliographystyle{aa}
\bibliography{biblio}

\begin{appendix}

\section{Searches for counterparts to \srcmm}
\label{sec:appendix}
We searched for a counterpart to \srcmm\ at other wavelengths using both simultaneous and archival observations of the field.
No UV emission was detected at the source position in the \swift/UVOT observations simultaneous with ALMA on the night of 2021 June 26--27, either in the images collected separately during the three snapshots or in the stack of these images.  We set a 3$\sigma$ upper limit on the magnitude of a putative counterpart of UVM2$>$21\,mag from the stacked image (total exposure of $\sim$2.4\,ks).
A few weeks before the ALMA observations were acquired, the FOV of \srcmm\ was also observed in the optical ($g'$, $r'$ and $i'$ SDSS filters) with the Las Cumbres Observatory (LCO) network of telescopes. No counterpart was detected in these data. We set the following 3$\sigma$ upper limits on the optical magnitude using 60 s images acquired with the 1m telescope of the LCO network located in South Africa on 2021 May 8: $i'>17.8$\,mag, $r'>17.9$\,mag, $g'>18.6$\,mag. There is also no optical/NIR counterpart in either the third data release (DR3) from the {\it Gaia} space telescope or in the 2MASS point source catalogue. \srcmm\ was also not detected in the VLA observation carried out on 2021 June 3, down to a 3$\sigma$ flux density upper limit of 12\,$\mu$Jy at 10\,GHz.

Several observations of the field have been performed in the soft X-ray band over the past years, mainly using \swift\ and \xmm. However, given the X-ray luminosity of \src\ and its small angular distance from \srcmm\ ($\sim$22\arc\ ), the point-spread function of the X-ray instruments on board these observatories encompasses the position of \srcmm, preventing a conclusive assessment of the presence or absence of an X-ray counterpart to this source.
We then analysed a 83-ks long \cxo/ACIS-S (Advanced CCD Imaging Spectrometer) observation performed in March 2011 (Obs. ID 11075) to search for X-ray emission at the position of \srcmm. We reprocessed the data using the task \textsc{chandra}$_-$\textsc{repro} within the \textsc{ciao} software and ran the \textsc{wavdetect} source detection tool. No evidence for an X-ray counterpart to \srcmm\ was found. Using \textsc{srcflux}, we set a 3$\sigma$ upper limit on the net count rate of $1.3\times10^{-4}$ counts\,s$^{-1}$ over the 0.5--7.0\,keV energy range. Assuming an absorbed power-law spectrum with an absorption column density of $N_{\rm H}$ = $5\times10^{20}$\,cm$^{-2}$ (i.e. the one expected within the Galaxy in the direction of the source: \citealt{willingale13}) and a photon index in the range $\Gamma$ = 1 -- 4, the limit above translates into an unabsorbed flux of $< (2-4)\times10^{-15}$ \flux\ (depending on the assumed power-law slope) in the 0.5--7\,keV range.
\end{appendix}

\end{document}